\documentclass[a4paper,11pt]{article}
\pdfoutput=1 

\usepackage{jinstpub} 

\usepackage{cleveref}
\crefname{figure}{figure}{figures}
\usepackage{caption}
\usepackage{subcaption}
\usepackage{alltt}
\usepackage{epstopdf}
\usepackage{setspace}
\usepackage{lineno}
\usepackage{dcolumn}
\usepackage[perpage]{footmisc}
\usepackage{amsmath,amsfonts,amssymb,wasysym,ifsym}
\usepackage{color}
\usepackage{MnSymbol}
\usepackage{verbatim}
\usepackage{upgreek}
\usepackage{hyperref}
\usepackage{bm}
\usepackage{textcomp}
\usepackage{graphicx}
\usepackage[T1]{fontenc}
\usepackage{libertine}
\usepackage{xspace}
\newcommand{\GEANTfour} {{\textsc{Geant4}}\xspace}

\title{SoLid: A short baseline reactor neutrino experiment}
\makeatletter
\newcommand\footnoteref[1]{\protected@xdef\@thefnmark{\ref{#1}}\@footnotemark , }
\makeatother

\collaboration{The SoLid Collaboration}
\author[a]{Y.~Abreu,}
\author[i]{Y.~Amhis,}
\author[b]{L.~Arnold,}
\author[g]{G.~Barber,}
\author[a]{W.~Beaumont,}
\author[n]{S.~Binet,}
\author[h]{I.~Bolognino,}
\author[i]{M.~Bongrand,}
\author[g]{J.~Borg,}
\author[i]{D.~Boursette,}
\author[d]{V.~Buridon,}
\author[j]{B.~C.~Castle,}
\author[n]{H.~Chanal,}
\author[b]{K.~Clark,}
\author[k]{B.~Coup\'e,}
\author[n]{P.~Crochet,}
\author[b]{D.~Cussans,}
\author[a,e]{A.~De Roeck,}
\author[d]{D.~Durand,}
\author[l]{T.~Durkin,}
\author[h]{M.~Fallot,}
\author[k]{L.~Ghys,}
\author[h]{L.~Giot,}
\author[g]{K.~Graves,}
\author[d]{B.~Guillon,}
\author[h]{D.~Henaff,}
\author[g]{B.~Hosseini,}
\author[i]{S.~Jenzer,}
\author[k]{S.~Kalcheva,}
\author[c]{L.N.~Kalousis,}
\author[f]{M.~Labare,}
\author[d]{G.~Lehaut,}
\author[b]{S.~Manley,}
\author[i]{L.~Manzanillas,}
\author[k]{J.~Mermans,}
\author[f]{I.~Michiels,}
\author[n]{S.~Monteil,}
\author[f,k]{C.~Moortgat,}
\author[b,l]{D.~Newbold,}
\author[d]{V.~Pestel,}
\author[b]{K.~Petridis,}
\author[a]{I.~Pi\~nera,}
\author[k]{L.~Popescu,}
\author[i]{N.~Roy,}
\author[f]{D.~Ryckbosch,}
\author[j]{N.~Ryder,}
\author[g]{D.~Saunders,}
\author[i]{M.-H.~Schune,}
\author[h]{M.~Settimo,}
\author[a]{H. Rejeb Sfar,}
\author[i,m]{L.~Simard,}
\author[g]{A.~Vacheret,}
\author[f]{G.~Vandierendonck,}
\author[k]{S.~Van Dyck,}
\author[c]{P.~Van Mulders,}
\author[a]{N.~van Remortel,}
\author[a,c]{S.~Vercaemer,}
\author[a]{M.~Verstraeten,}
\author[h]{B.~Viaud,}
\author[j,l]{A.~Weber}
\author[h]{and F.~Yermia.}

\affiliation[a]{Universiteit Antwerpen, Antwerpen, Belgium}
\affiliation[b]{University of Bristol, Bristol, UK}
\affiliation[c]{Vrije Universiteit Brussel, Brussel, Belgium}
\affiliation[d]{Normandie Univ, ENSICAEN, UNICAEN, CNRS/IN2P3, LPC Caen, 14000 Caen, France}
\affiliation[e]{CERN, 1211 Geneva 23, Switzerland}
\affiliation[f]{Universiteit Gent, Gent, Belgium}
\affiliation[g]{Imperial College London, Department of Physics, London, United Kingdom}
\affiliation[h]{SUBATECH, CNRS/IN2P3, Universit\'e de Nantes, Ecole des Mines de Nantes, Nantes, France}
\affiliation[i]{LAL, Univ Paris-Sud, CNRS/IN2P3, Universit\'e Paris-Saclay, Orsay, France}
\affiliation[j]{University of Oxford, Oxford, UK}
\affiliation[k]{SCK-CEN, Belgian Nuclear Research Centre, Mol, Belgium}
\affiliation[l]{STFC, Rutherford Appleton Laboratory, Harwell Oxford, and Daresbury Laboratory, Warrington, United Kingdom}
\affiliation[m]{Institut Universitaire de France, F-75005 Paris, France}
\affiliation[n]{Universit\'e Clermont Auvergne, CNRS/IN2P3, LPC, Clermont-Ferrand, France}

\emailAdd{nick.vanremortel@uantwerpen.be, guillon@in2p3.fr}

\abstract{The SoLid experiment, short for Search for Oscillations with a Lithium-6 detector, is a new generation neutrino experiment which tries to address the key challenges for high precision  reactor neutrino measurements at very short distances from a reactor core and with little or no overburden. The primary goal of the SoLid experiment is to perform a precise measurement of the electron antineutrino energy spectrum and flux and to search for very short distance neutrino oscillations as a probe of eV-scale sterile neutrinos. This paper describes the SoLid detection principle, the mechanical design and the construction of the detector. It then reports on the installation and commissioning on site near the BR2 reactor, Belgium, and finally highlights its performance in terms of detector response and calibration.{ \color{red} \it \footnotesize  }
}

\begin{document}
\maketitle
\clearpage
\section{Introduction}\label{sec:intro}

Sterile neutrinos, originally introduced by Bruno Pontecorvo in 1967 \cite{Pontecorvo_1968}, are well-motivated in many extensions of the Standard Model as they appear in most of the possible mechanisms to explain neutrino masses. Apart from these theoretical considerations, the first hints appeared from accelerator-based neutrino experiments, LSND then MiniBoone, which have observed persistent anomalies, in electron neutrino appearance and muon neutrino disappearance~\cite{ref:LNSD,ref:MiniBoone}. The second hints arose from solar neutrino experiments, SAGE and GALLEX, which measured a significant deficit of the neutrino flux, when using high-activity $\nu_e$ sources during calibration runs~\cite{ref:Gallium1,ref:Gallium2,ref:KOSTENSALO2019542}. The third indication came in 2011 from the re-evaluation of the $\bar \nu_e$ reactor flux obtained with a state-of-the-art prediction model. It exhibits a 6\% average deficit on the measured antineutrino flux~\cite{PhysRevC.83.054615,Mention_2011, PhysRevC.84.024617}. This deficit, known as the Reactor Antineutrino Anomaly (RAA), is significant at the 2.5 $\sigma$ level.
Though some tensions persist when combining both LSND and MiniBoone results with reactor measurements, no phenomenological models are known to better fit all the data than those adding sterile neutrinos at a mass scale of order 1~eV$^2$~\cite{Dentler_2018,ref:Berryman}. The search for such a sterile neutrino  provides a clear motivation to measure the neutrino fluxes and spectra with dedicated experiments at very short baselines near nuclear reactors~\cite{abazajian2012light}. Several experiments world-wide have taken, or are taking physics data \cite{Ko:2016owz,Danilov:2013caa, Serebrov:2018oyd, Allemandou:2018vwb,  Ashenfelter_2019}. Some of these experiments already published constraints on the RAA allowed parameter space, that exclude the RAA best-fit point \cite{Mention_2011} at more than 95\% C.L.~\cite{PhysRevD.102.052002, andriamirado2020improved}. However, global fit analysis and others published results favour sterile neutrino oscillations at the 3 $\sigma$ level~\cite{serebrov2020analysis, GARIAZZO201813, Dentler_2018}. In addition to reactor experiments, an active search is also performed by using accelerator decay-in-flight neutrino beams. The US is currently running the Booster Neutrino Beam (BNB) which will enable the deployment of multiple detectors at different baselines: SBND, MicroBooNE and ICARUS~\cite{ref:BNB}.~\\

Besides the eV-scale neutrino search, recent $\theta_{13}$  precision measurements are all indicating a deviation in the $\bar \nu_e$ energy spectrum shape, between 5 and 6 MeV~\cite{Abe:2014bwa,PhysRevLett.116.061801,RENO:2015ksa}, also known as the ’5 MeV bump’. 
It is likely related to nuclear and reactor physics and thus puts the $\bar \nu_e$ flux prediction and its uncertainties estimation into question \cite{PhysRevLett.119.112501}. Nuclear models are scrutinized and many dependencies are currently investigated: fission yields and their dependencies with neutron energy spectrum, beta spectrum shape and weak magnetism correction \cite{PhysRevC.95.064313,PhysRevC.100.054323}, time-dependent relative contribution of fissile isotopes ($^{235} \rm U$, $^{239} \rm Pu$) \cite{PhysRevLett.120.022503} or the energy response linearity of detectors \cite{MENTION2017307}. In this context, the neutrino energy spectrum distortion is currently being investigated by very short baseline experiments near HEU (Highly Enriched Uranium) research reactor \cite{andriamirado2020improved, stereocollaboration2020accurate}.\\

SoLid, or Search for oscillations with a Lithium 6 detector, is a very short baseline neutrino oscillation 
experiment, located near the BR2 reactor of the SCK\raisebox{-0.8ex}{\scalebox{2.8}{$\cdot$}}CEN 
in Belgium. Its main purpose is to perform a precise measurement of the electron antineutrino 
energy spectrum and flux as a function of the distance travelled by antineutrinos between the reactor 
core and their interaction in the detector. These measurements will be primarily used to search 
for the existence of one or more sterile neutrinos corresponding to mass eigenstates of order 
$\Delta m^2\sim 1\,$eV$^2$. Secondarily, the shape of the energy spectrum will serve as a reference 
measurement for electron antineutrinos originating from the fission of~$^{235}$U. In order to 
achieve these goals, the SoLid experiment aims to detect electron antineutrinos with a target efficiency 
of at least 10\%, reconstruct their energy with a resolution of 14\%/$\sqrt{E\text{(MeV)}}$, and 
obtain an overall Signal to Background ratio (S/B) of order unity, given that it operates with a minimal overburden of only 8~m water equivalent.\\

Operating very close to the reactor core and at sea level, where large cosmic and reactor backgrounds are 
produced, combined with small installation spaces, represents several challenges in terms of background 
rejection capabilities. Compared to the contemporary very-short baseline neutrino experiments near 
reactors~\cite{Ko:2016owz,Danilov:2013caa, Serebrov:2018oyd, Allemandou:2018vwb, Ashenfelter:2018cli}, 
the SoLid detector has some unique features, which are described extensively in~\cite{Abreu:2017bpe}. It uses 
a finely 3D segmented plastic scintillator to detect electromagnetic energy deposits, combined with scintillation screens that contain $^6$Li that provide distinct nuclear induced signals. The use of high segmentation and the dual scintillator, provides particle discrimination, and aims to identify and reduce backgrounds. Moreover, the materials used, the robustness and compactness are also attractive for future reactor monitoring applications.\\

 After demonstrating the applicability of the composite scintillator technology, a full-scale prototype module, SM1, with a fiducial mass of 288$\,$kg was operated near the BR2 reactor in 2015. Based on the performance 
of the prototype module~\cite{Abreu:2018pxg}, improvements were made to the original detector design 
before proceeding to the construction of a 1.6 ton detector in 2016-2017. The SoLid detector installation 
was completed in February 2018 and was successfully commissioned near the BR2 reactor in spring 2018. 
In this paper, we will first give a complete description of the SoLid detector: its detection principle, its mechanical design, the construction phase and the quality assurance process. We will then describe the dedicated front-end electronics and the data acquisition system. In the third part, we will present the BR2 reactor core near which the SoLid experiment operates. The BR2 reactor has very little neutron and gamma background, due to its moderate thermal power, adequate shielding, and absence of other experiments in the vicinity of our experiment. Its core size is also very compact. Finally, we will present the data taking operation and describe how the detector response is simulated and how the experiment is calibrated in-situ. The SoLid experiment has collected data up to june 2020. However, this paper describes the detector commissioning, calibration and measurement stability for the first two years of data taking, covering the period July 2018 - August 2019.
%

\section{Detector layout and design}\label{sec:layout}

 \subsection{Detection principle}\label{sec:det_principle}

The SoLid detector is designed to be a highly 3D segmented detector (8000\,voxels/m$^3$) based on a dual 
scintillation technology. Electron antineutrinos will interact primarily in the active detector volume via inverse beta 
decay (IBD) on hydrogen nuclei, producing a positron and a neutron in the final state: $\bar \nu_e + \text{p} \rightarrow e^+ + \text{n}$. Experimental approaches use the coincidence technique, which consists of detecting both the positron and the neutron, within a short time window, typically up to hundreds of microseconds \cite{ref:Cowan}. The neutron generally thermalizes via elastic collisions in the detector, 
after which it can be captured by nuclei with a high neutron capture cross section. As such it typically induces a 
scintillation signal that is delayed in time with respect to the scintillation light caused by the 
positron and its corresponding annihilation gamma-ray photons. The time delay between the two signals can 
be tuned by the choice of neutron capture elements and their concentration and distribution in 
the detector. Neutrino oscillation experiments typically vary in their choice of scintillator, the neutron 
capture element, and the way these are incorporated in the scintillator~\cite{ref:Bugey3,ref:JUNO,ref:Borexino,ref:Kamland,ref:DayaBay,ref:doublechooz,ref:reno,ref:Angra}. \\

SoLid opted for a combination of two scintillators. One is polyvinyl toluene (PVT), a relatively cheap plastic 
scintillator that is generally easy to machine in any desired shape or geometry, and the other 
is ZnS(Ag) used together with $^6$LiF to capture thermal neutrons via the reaction:
\begin{equation}\label{eq:LiCap}
^{6}_{3}\text{Li} + \text{n} \rightarrow ~ ^{3}_{1}\text{H} + \alpha~\text{(4.78\,MeV)},
\end{equation}
for which the decay products in turn induces scintillation in the ZnS(Ag) scintillator. The PVT-based scintillator is of 
the type EJ-200 produced by ELJEN Technology. It is a general purpose plastic scintillator that emits on average 
10 000 photons per MeV of energy deposited by electrons of 1\,MeV in the blue-violet wavelength band with a 
peak emission wavelength of 425\,nm~\cite{ref:ELJEN}. The choice of PVT is mainly motivated by its good light yield and its linear 
energy response over a wide range of energies ranging from 100\,keV to several MeV. It combines a long optical attenuation length of about 380 cm, with a scintillation pulse decay time of 2.1\,ns. The $^6$LiF:ZnS(Ag) scintillator for neutron detection is produced by SCINTACOR, in the form of thin screens~\cite{ref:Scintacor}. These so-called neutron detection screens, emit photons at a peak emission wavelength of 450\,nm. The nature of the neutron capture reaction and the longer scintillation decay time of 10 microseconds for the $^6$LiF:ZnS(Ag) scintillator allows for a pulse shape discrimination between signals induced in the neutron detection screens via nuclear interaction, hereafter denoted as {\em NS}, and signals induced via electromagnetic processes in the PVT, denoted as {\em ES}.\\

The detection technology, the materials and the geometrical arrangement of the main components of the SoLid detector are the same as for the SM1 prototype and are outlined in an earlier paper~\cite{Abreu:2018pxg}. Based on the performance of this full-scale prototype and on studies with a dedicated test bench, described in~\cite{Abreu:2018ajc}, several design improvements were made to optimize the uniformity of the detector response and to maximise the light collection efficiency. These improvements are outlined below.

 \subsection{Mechanical design}\label{sec:design}

 \subsubsection{Detection cell}\label{sec:det_cell}

The basic detection cell consists of a 5x5x5\,cm$^3$ PVT cube, of which two faces are covered with neutron detection screens. Positrons with an energy of 10\,MeV travel less than 48\,mm in PVT, which implies that the majority of the IBD positrons will be stopped in the same cell as in which they are produced. In order to extract the scintillation photons produced in the PVT or in the neutron detection screens, 4~grooves with a 5$\times$5\,mm$^2$ square cross section are machined in four different faces of each cube. Each groove accommodates an optical fibre with a square cross section of  3$\times$3\,mm$^2$ that guides the light to an optical sensor at the edge of the detector. All detection cells are optically isolated via a DuPont Tyvek wrapping of type 1082D~\cite{ref:Tyvek}, whose thickness has been increased from 205 to 270\,\textmu{}m to reduce the optical transparency.
\begin{figure}[h!]
	\centering
	\includegraphics[width=1.\textwidth]{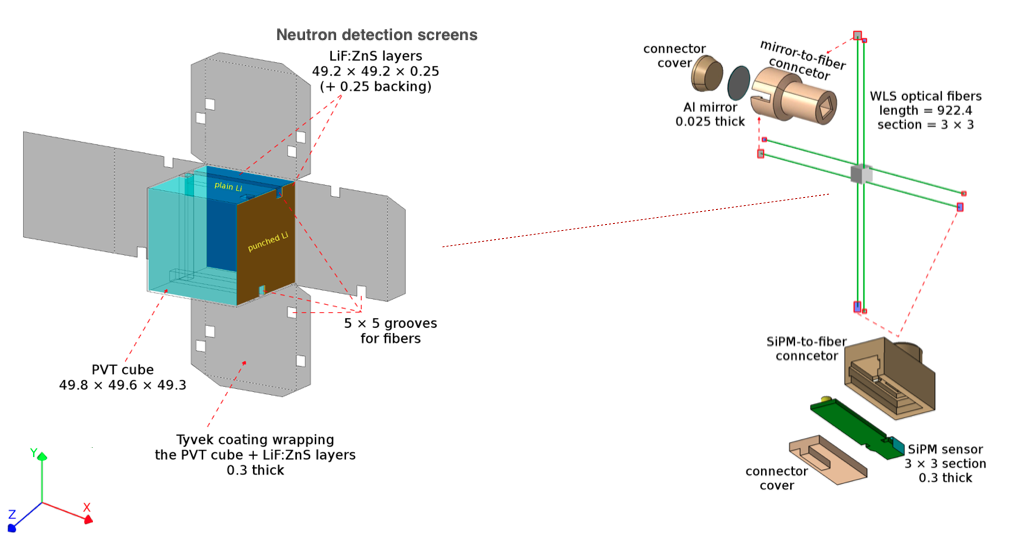}
	\caption{{\small{(Left) A schematic view of the PVT detection cell, including the two, asymmetrically placed, neutron detection 
	screens and its Tyvek wrapping. All cells are arranged such that the face labeled as "plain Li" (dark blue) 
	faces the BR2 reactor. (Right) Four wavelength shifting fibres cross each detection cell, with alternating positions 
	of MPPCs and mirrors at the fibre ends. They are contained in plastic printed connectors. Dimensions are in mm.}}}
	\label{fig:phase1_cube}
\end{figure}

The neutron detection screens are cut into squares of 5$\times$5\,cm$^2$ and positioned, using no glues or optical gels, on two adjacent faces of the PVT cube. The two cube faces that are covered with neutron detection screens are the one that faces the reactor core, perpendicular to the Z-axis, and the one that is perpendicular to the X-axis, 
facing the electronic readout boxes that are mounted on one detector side. A schematic view of a detection cell together with the coordinate system and the position of the neutron detection screens is shown in Fig.\,\ref{fig:phase1_cube}. The scintillation light produced in the neutron detection screens is optically coupled to  the PVT cube via the air trapped in between the two surfaces. The bulk of the neutron detection screens have a 225\,\textmu{}m thick MELINEX-339 reflective backing. The addition of this backing on the neutron detection screens with respect to the prototype module, combined with the overall improved light detection in the cells increases the amplitude 
of the {\em NS} signals and improves the {\em NS}-{\em ES} waveform discrimination. By doubling the amount 
of neutron detection screens per cell, and due to the asymmetric placement around each cube, the capture efficiency for thermal neutrons in the SoLid detector was optimized and increased by 30\%, compared to the SM1 prototype~\cite{Abreu:2018pxg}. The capture time is also reduced from 102 to 65 microseconds, which decreases the background from random coincidences in the offline analyses.\\

The LiF component of the neutron detection screens is the dominant component in terms of radiopurity. Bulk amounts of this material were measured in the underground low background radiation facilities of Modane and Boulby. The most accurate measurement on the activity of the LiF yielded an activity of 69$\pm$35 mBq/kg of pure LiF. The second largest contamination is ZnS, for which the upper limit on the rate is at least 5 times smaller. Our detector has a total of 8.9 kg of LiF, which yields a rate of 614$\pm$ 311 mBq, which is also consistent with the measured intrinsic background rate by a dedicated analysis.

 \subsubsection{Light collection}\label{sec:light_collect}

%
The scintillation photons produced in each detection cell are extracted and guided by 92~cm long 
double clad wavelength shifting fibres (494 nm), of type BCF-91A, produced by St.Gobain~\cite{ref:SaintGobain}. One end of each optical fibre 
is covered by a Mylar foil with a reflective aluminium coating, and the other end is coupled to a 
Hamamatsu type S12572-050P multi-pixel photon counter (MPPC), containing~3600~pixels, 
arranged in a 3$\times$3~mm$^2$ matrix ~\cite{ref:MPPC}. For our current settings, the photon detection efficiency is 32\%. The position of the MPPC and mirror alternates between 
the parallel fibres to mitigate the attenuation of light in the fibres and to ensure a more uniform 
light response throughout the detector (see Fig.\,\ref{fig:phase1_cube}). The detection 
cells are arranged into a detection plane of 16$\times$16 cells, where each row and column of cells is 
read out by the same set of two optical fibres, accounting for a total of 64 optical fibres, and an equal number of readout channels, per detector plane, 
as shown in Fig.\,\ref{fig:phase1_module}.

 \subsubsection{Plane \& module design}\label{sec:plane_module}

\begin{figure}[h!]
	\centering
	\includegraphics[width=1.\textwidth]{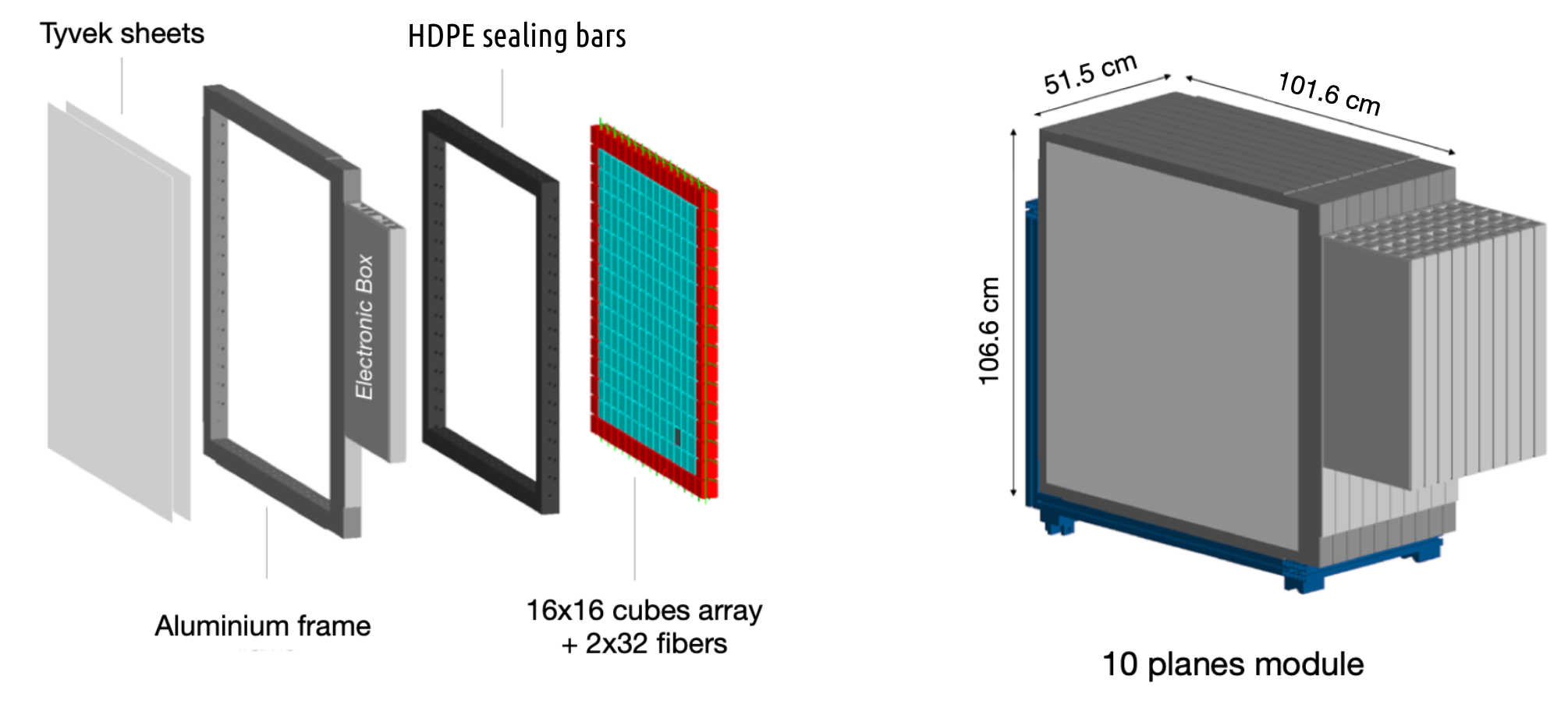}
	\caption{{\small{(Left) Exploded view of a detection plane, showing the 16x16 detection cells (blue and red). The outer layer of active cells have neutron screens without a reflective backing (red). The active volume is surrounded by the HDPE neutron reflector (black),the aluminum frame for mechanical rigidity and attachment of the electronics, and two tyvek sheets for light isolation. (Right) Sketch and dimensions of a 
	10 planes detector module mounted on its trolley (blue).}}}
	\label{fig:phase1_module}
\end{figure}

%
The detection planes, with a cross sectional surface of 0.8$\times$0.8\,m$^2$, are surrounded 
by a lining of white high-density polyethylene (HDPE) with a thickness of 46.0 and 46.8\,mm, 
respectively in the vertical and horizontal directions (see Fig. \ref{fig:phase1_module}). The 
HDPE bars act as reflectors for neutrons that would otherwise escape the detector. Each plane 
is structurally supported by a hollow frame of extruded aluminium that has been chrome coated 
to act as a Faraday cage for the MPPCs and their wirings. Each fibre protrudes through the HDPE 
lining and the frame where it is capped off on each end with two different plastic 3D printed caps. 
One cap holds an MPPC sensor, while the other end holds the aluminized Mylar mirror (see Fig.~\ref{fig:phase1_cube}). Optical contact with both the mirror and the MPPC is ensured with a drop of optical gel. The MPPC 
bias voltage and signal is carried on twisted pair ribbon cables that are routed through the hollow frame and 
are terminated on one of the frame sides in four insulation displacement connectors (IDCs) each grouping 16 MPPC 
channels. The front-end electronics, which is described in section~\ref{sec:daq}, is self-contained in an 
aluminium encasing mounted on one side of each detection plane. Each detection plane is finally 
covered with two square Tyvek sheets on each of its light sensitive faces to further ensure optical isolation 
from its neighbouring planes.\\ 

Frames and their attached readout electronics are grouped together by 10 units to form 
a detector module, mounted on a trolley (see Fig. \ref{fig:phase1_module}). Each 
module can be operated as a standalone detector and has its own power supply and trigger 
electronics mounted on an overhead rail (see section \ref{sec:daq}). The 
SoLid detector currently includes a total of 5 detector modules, accounting for a total of 50 detector planes 
and corresponding to a fiducial mass of 1.6\,ton. The front and  back planes of the detector are capped with 
a HDPE reflective shielding with a thickness of 9\,cm. Under normal detector operations all modules are 
closely grouped together with an average spacing of 0.5\,mm between two modules.

 \subsection{Detector construction}\label{sec:det_construction}

 \subsubsection{Cell production and assembly}\label{sec:cube_production}
 
\begin{figure}[h!]
	\centering
	\includegraphics[width=.95\textwidth]{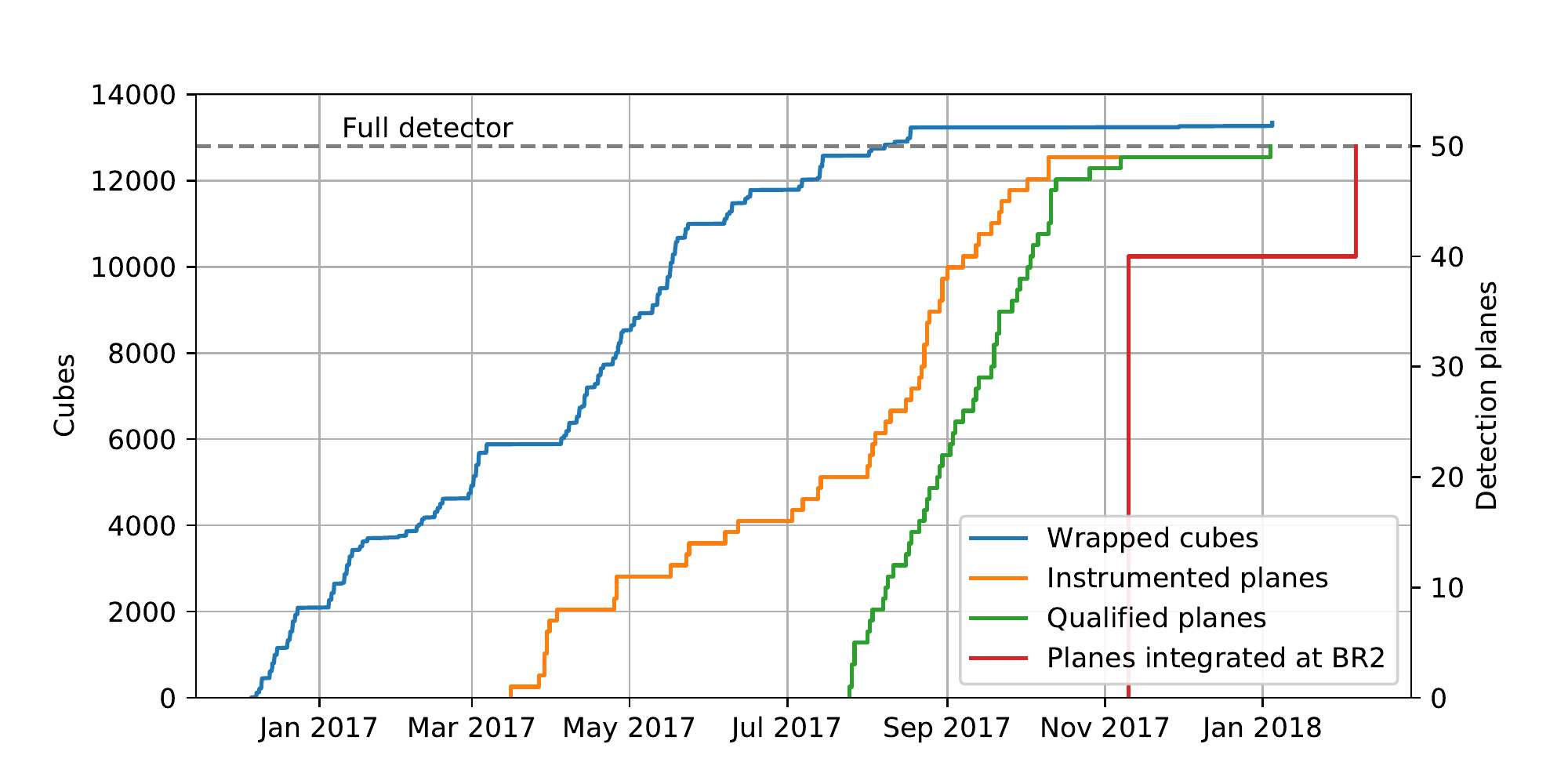}
	  \vspace*{-0.2cm}
	\caption{{\small{Time evolution of the SoLid detector construction phase, illustrating the progress of each of the main construction phases: cube production, frame filling and cabling, quality testing and integration. }}}
	\label{fig:assembly}
\end{figure}

The construction of the SoLid detector started in December 2016 and took roughly 14 months. 
The progress of the detection cells (wrapped cubes) production and plane assembly is shown in Fig.\,\ref{fig:assembly}. The PVT cubes were extracted from 104$\times$52$\times$6.3\,cm$^3$ PVT slabs and individually machined by an industrial partner in Flanders using CNC milling machines, with 0.2\,mm tolerance on the cube and groove dimensions. After milling, all cubes were visually inspected for mechanical damage before being transported 
to the integration site at Universiteit Gent. There all cubes were washed with a light soap detergent to remove 
lubricant from the milling process and dried overnight. During frame production, two types of neutron detection screens were used. The cells contained in the bulk of the detector are all equipped with neutron detection screens that have a backing with a thickness of 225\,\textmu{}m, while all cells located at the outer edge of each frame received neutron detection screens without reflective backing material, that were left over from the construction of the SM1 prototype.\\ 

\begin{figure}[h!]
	\centering
         \includegraphics[width=1.\textwidth]{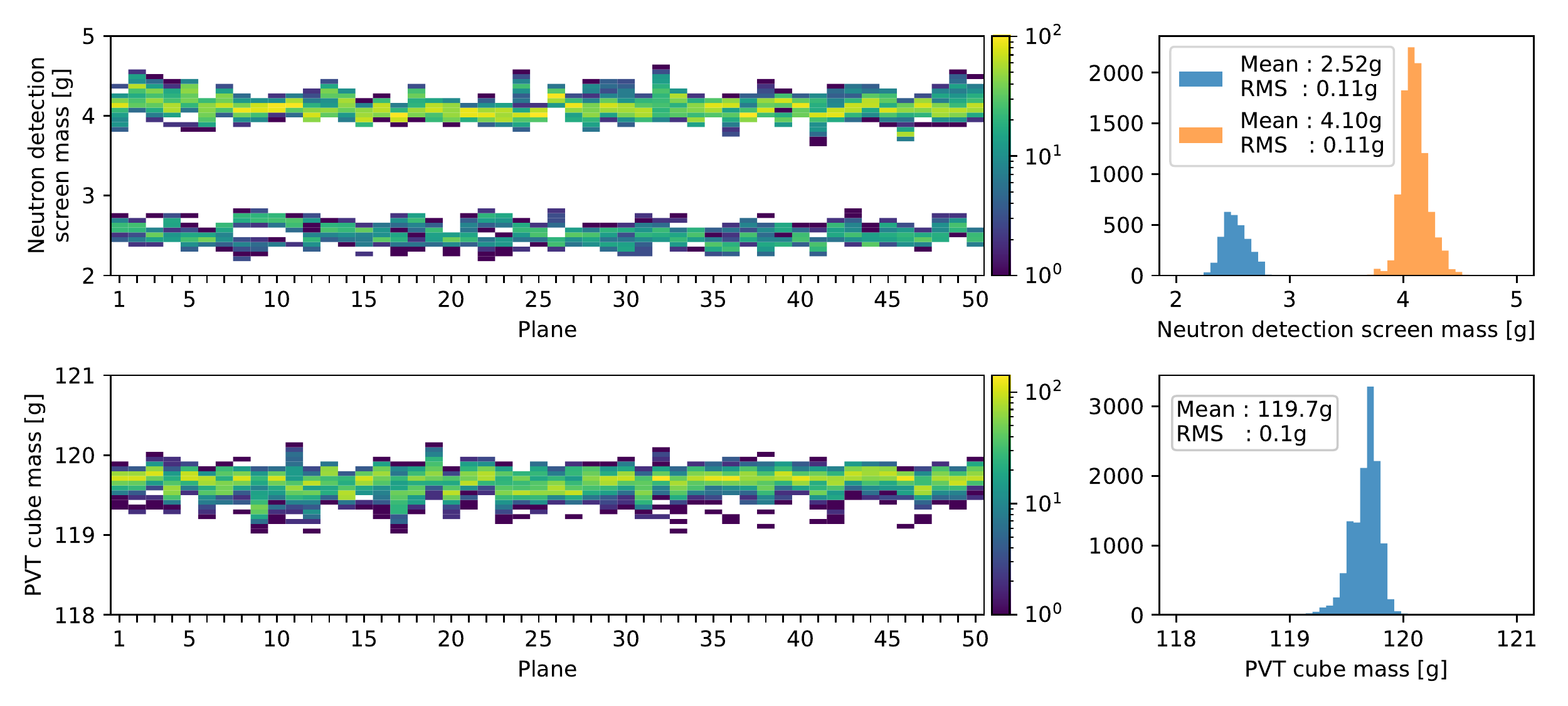}
         \vspace*{-0.6cm}
	\caption{{\small{Distribution of the masses for the 2 types of neutron detection screens (top) and for the PVT cubes (bottom), along each of the detector planes (left), and distributed throughout the whole detector (right). }}}
	\label{fig:Mass_dist}
\end{figure}

Each cube was weighed with a digital scale with a precision of 1\,mg, before and after being 
equipped with neutron detection screens and wrapped with Tyvek. The two neutron detection screens 
for each detection cell were also individually weighted. Each detection cell was marked with a bar code 
sticker that allows for tracking of the production history in a dedicated SQL database. This database includes 
the bare and wrapped weights of each cell. During a period of 8 months a total of 13228 cubes were washed,
 inspected, wrapped and catalogued. Only 3\% of all produced PVT cubes 
 were rejected due to quality issues. The accuracy of the weights, combined with the tracking of the 
 production batches revealed a small shift in cell mass during the production process, which 
 falls well within the tolerances used in the cell quality control. The mass distributions 
of the PVT and neutron detection screens of the 50 detection planes are shown in Fig.\,\ref{fig:Mass_dist}. 
The mean weight of all PVT cubes equals 119.7\,g with an RMS of~0.1\,g, which allows to control at per mille 
level the proton content. The difference in mass between the neutron detection screens with and without 
reflective backing can be observed in Fig.\,\ref{fig:Mass_dist}. Each of the 50 detection planes was assembled and equipped by hand in its 
aluminium frame. The position of each MPPC in the detector is stored in the construction database, together 
with its breakdown voltage. 
%

 \subsection{Quality assurance}\label{sec:quality}

\begin{figure}[h!]
	\centering
	\includegraphics[width=.5\textwidth]{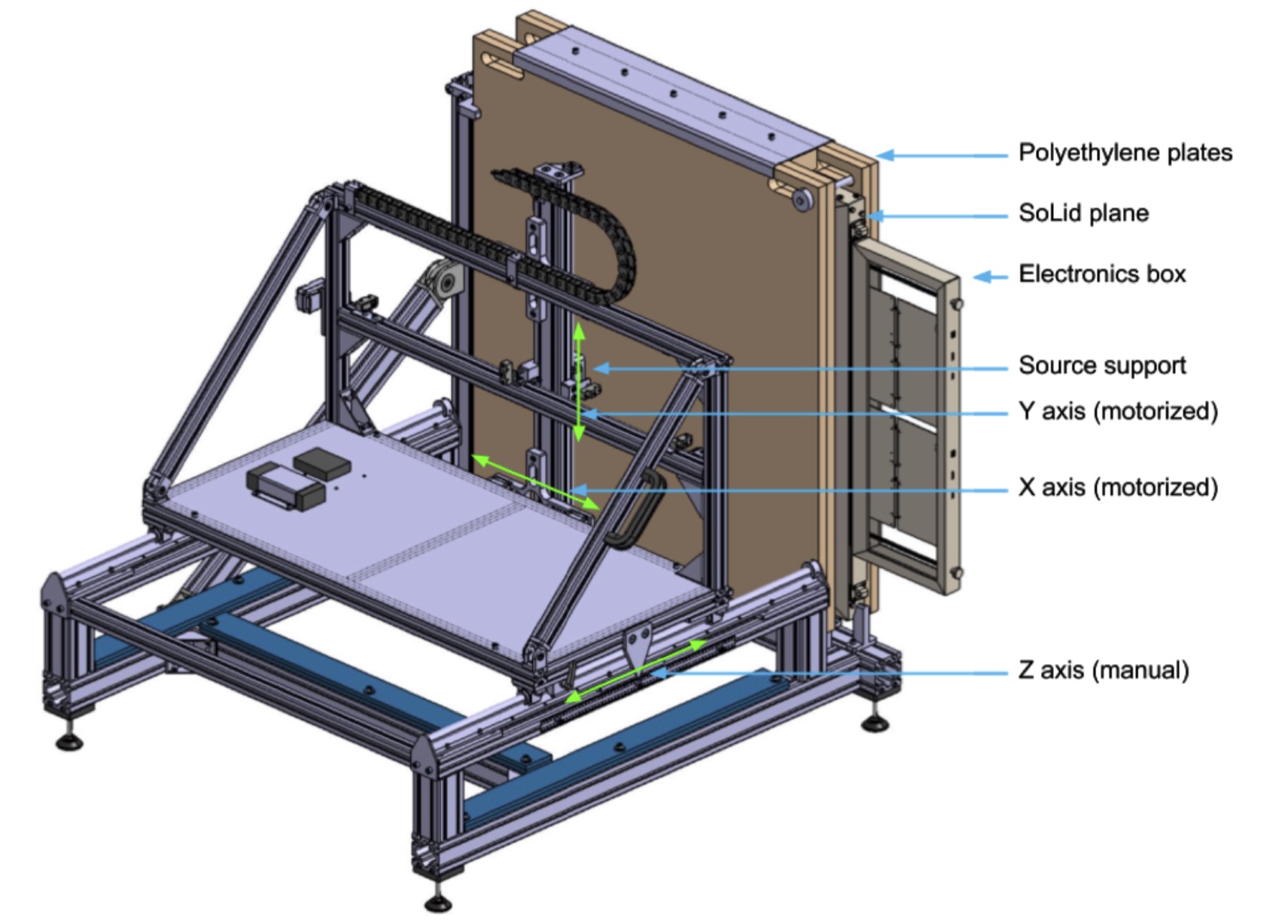}
	\includegraphics[width=.48\textwidth]{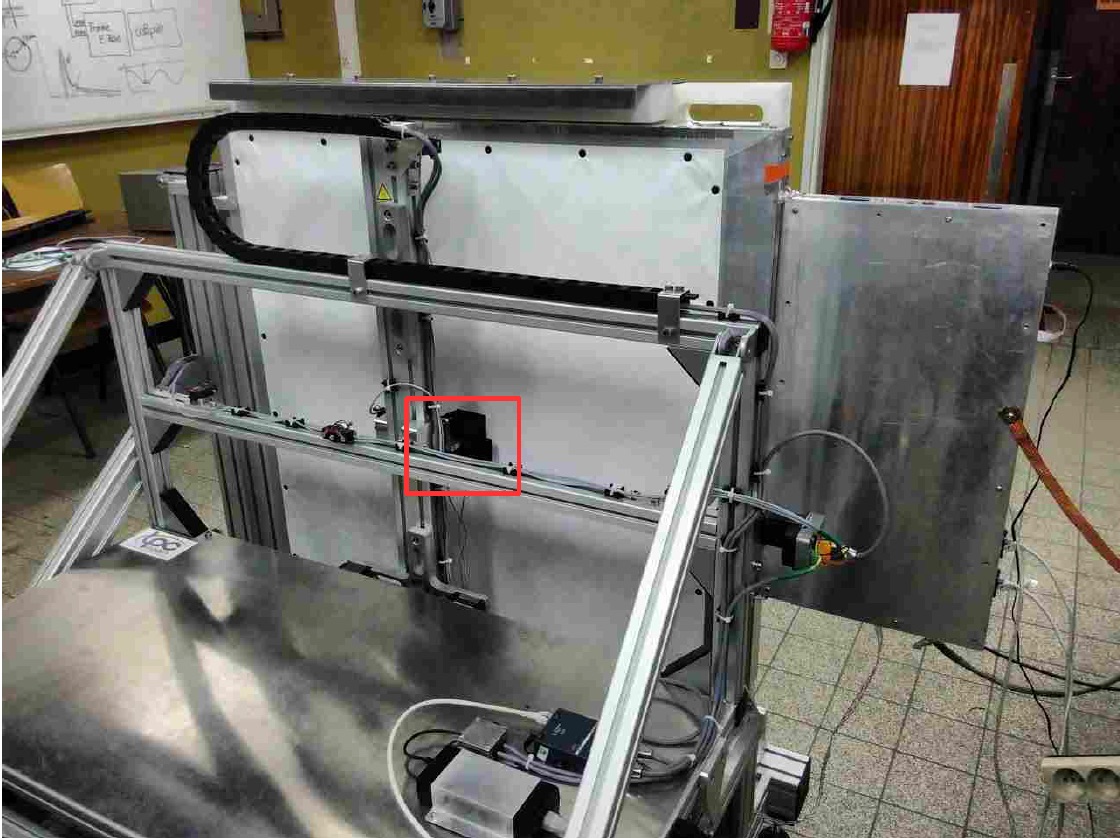}
	\caption{{\small{(Left) Mechanical design of the Calipso test bench, showing a detector plane under test. (Right) Photograph of 
	Calipso with a SoLid detection plane to be calibrated. The source is located in the black 
	box in the red square. The electronic box is on the right of the plane.}}}
	\label{fig:calipso}
\end{figure}

Before being integrated in a detection module, each detection plane was tested on the so-called Calipso test bench, shown in Fig.\,\ref{fig:calipso} and described in detail in~\cite{Abreu:2018ekw}. This test bench consists of a robot that can position a calibration source in front of a SoLid plane with millimetre accuracy. A polyethylene (PE) neutron collimator is added when performing neutron calibrations, in order to increase the neutron capture rate. In addition, a dedicated $^{22}$Na self-triggering calibration head was designed for the calibration of the energy response of the PVT. The Calipso test bench served primarily as an automated quality control system. As such it provided an early detection of typical construction quality issues such as missing neutron detection screens, bad fibre connections, malfunctioning MPPCs and wrong cabling which were all resolved before integration in a detector module. It also allowed to perform an initial test of the electronics and DAQ system before mass production. As a result, for a nominal bias of 1.5\,V above each MPPC breakdown voltage, an average gain of about 22 Analogue-to-Digital Conversion units (ADC) per pixel avalanche (PA) was determined with an RMS of 3\%. This was further refined with in-situ equalizations during detector commissioning at the reactor site to achieve a gain equalized to 1.4\% across the whole detector (see section \ref{sec:monitoring}).\\

The quality assurance campaign with Calipso allowed to have a preliminary calibration of all the detection cells. Calipso measured the light yield by using a $^{22}$Na gamma source in coincidence with an external trigger 
to remove background. The measured Compton edges caused by the interaction of the 1270\,keV 
gamma rays are used to extract the light yield using two consistent methods based on an analytical 
fit and a template method described in~\cite{Abreu:2018ekw}. The average light yield was observed
 to be larger than 70\,PA/MeV/cell corresponding to a stochastic energy resolution of 12\% which is consistent with the SoLid physics requirements~\cite{Abreu:2018ajc}. Detailed in-situ calibrations are described in section~\ref{sec:calibration}, and result in a higher and more accurate light yield determination. The response of the detector to neutrons was also evaluated using a $^{252}$Cf source emitting neutrons with a mean energy of~2\,MeV in order to determine the relative difference in neutron response across the detector and to validate the neutron trigger settings as described in~\cite{Abreu:2018njy}. Because of the dependence on moderation and detector geometry, the absolute neutron capture and reconstruction efficiency is determined in-situ, as will be detailed in the section \ref{sec:calibration}. Combining the capture and reconstruction efficiency, the total relative dispersion of this efficiency across the detection cells is 5\%.

 \subsection{Container integration}\label{sec:container}

\begin{figure}[h!]
	\centering
	\includegraphics[width=.9\textwidth]{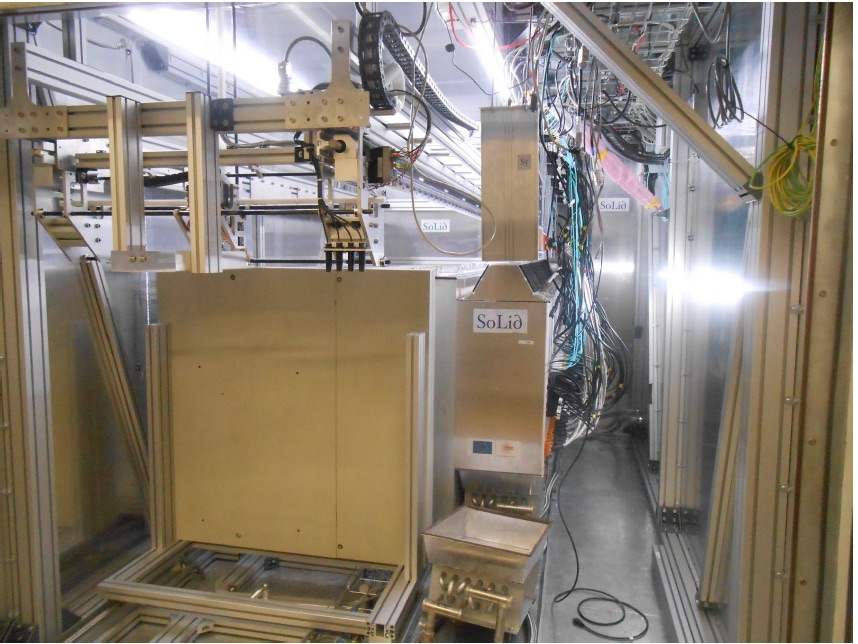}
	\caption{{\small{Photograph of the SoLid detector in its cargo container prior to the 
	installation of the last detection module. Environmental sensors are placed throughout the container, as well as Rn and gamma background monitors, which are mounted on the inner wall of the container on the right hand side of the detector.}}}
	\label{fig:phase1_photo}
\end{figure}

The detector and its electronics are installed in a cooled cargo container with dimensions of 2.4$\times$2.6$\times$3.8\,m$^3$ as shown in Fig.\,\ref{fig:phase1_photo} and \ref{fig:cross}. The container is further customized for thermal insulation and feed through of cooling lines. A dedicated patch panel, located on the side of the container, bundles all the connectors needed for the electronics (power supply, readout), the container instrumentation and the ethernet communication. The 5 detector modules are positioned off-center in the container in order to allow for access and service space (see~Fig.\,\ref{fig:phase1_photo}, \ref{fig:cross} and~\ref{fig:phase1_projected}). They are mounted on a rail system, that allows for an accurate and robust positioning and alignment (see CROSS calibration system in section \ref{sec:cross}).  The electronics are cooled by a chiller system which is described later in section \ref{sec:readout_design}. Due to the dimensioning of the chiller system and its radiators it is possible to cool down and control the ambient air temperature in the container to a precision of 0.2\,degrees Celsius. Under normal data taking circumstances, the ambient temperature of the SoLid detector is kept at a fixed value of 11\,degrees Celsius. In order to keep the relative humidity of the air inside the detector at acceptable levels the container is permanently flushed with dry air that enters the container at a low flow rate of 5\,m$^3$/hour. This flushing also helps to remove possible traces of Rn gas inside the detector. Environmental parameters such as pressure, 
 temperature and humidity in the container are constantly monitored by means of a custom sensor network that is controlled and read out by a Raspberry-Pi device. This specific readout is interfaced with the data acquisition of the experiment. During nominal data taking, the gamma background is monitored by a standard PMT coupled NaI scintillator, located inside the container. The airborne radon concentration is monitored by a radon detector, placed next to the NaI detector inside the container. The Rn measurement is performed by sampling the air with a small pump and sending it to a pin-diode semiconductor detector based on the RADONLITE and RADONPIX technology~\cite{Caresana_2014}, developed at CERN.

%

%
 \subsection{CROSS calibration system}\label{sec:cross}
%

\begin{figure}[h!]
\centering
	\includegraphics[width=.51\textwidth]{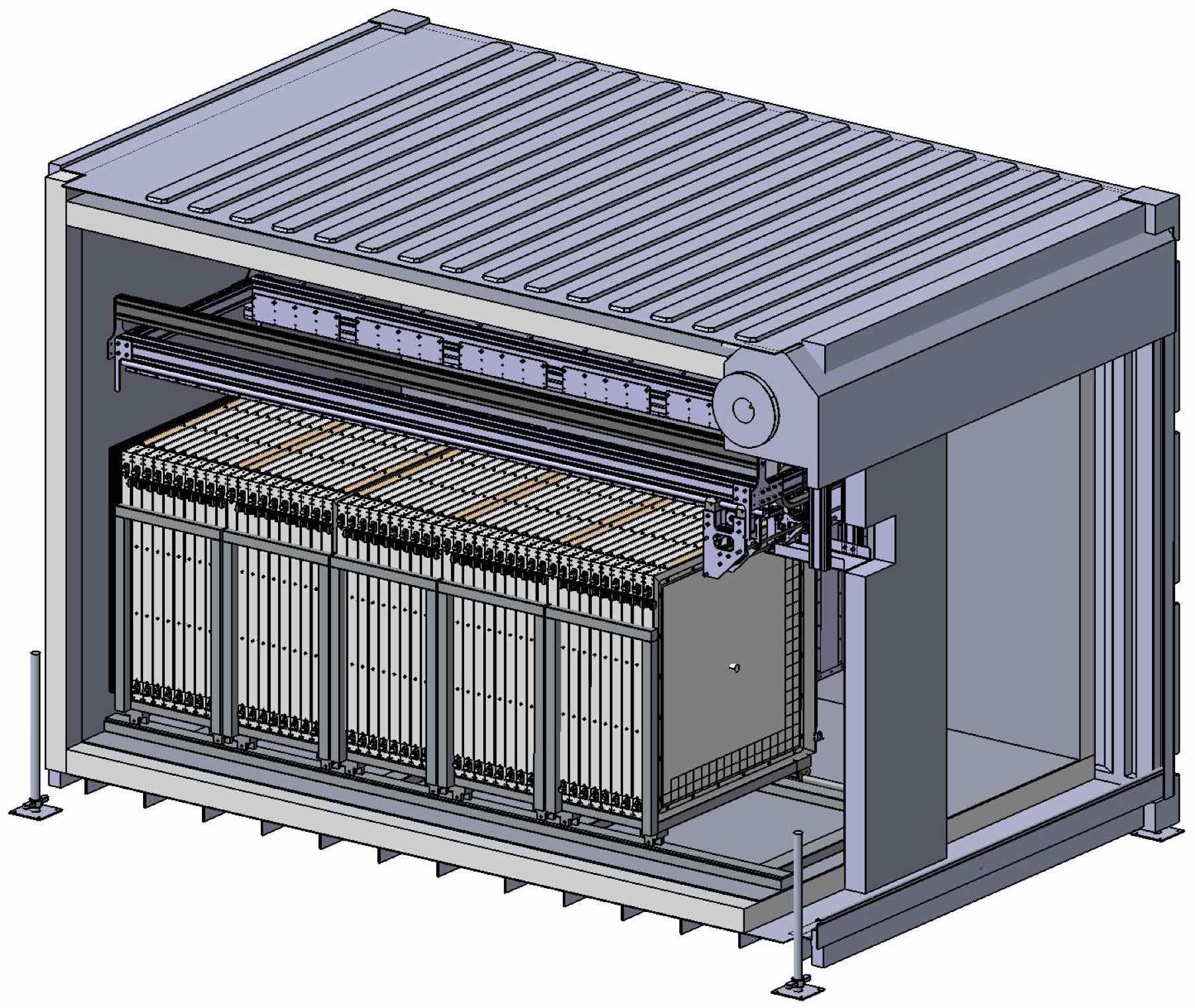}\hspace*{.3cm}~~~
	\includegraphics[width=.45\textwidth]{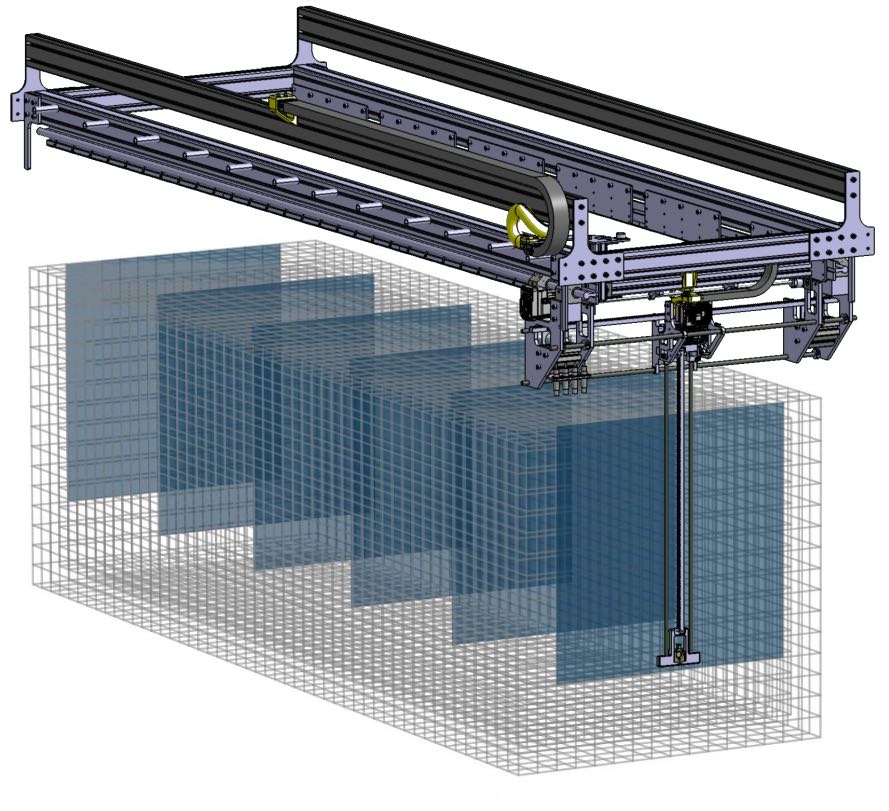}
	\caption{{\small{(Left) Sketch of the CROSS calibration robot and its ground rail system inside the container. (Right) 
	Sketch of the radioactive source holder and the area it can access within an open gap  indicated by the blue squares.}}}
	\label{fig:cross}
\end{figure}

In order to perform in-situ calibrations of the electromagnetic energy response and of the neutron 
capture efficiency, a calibration robot, CROSS, is mounted on top of the SoLid detector inside the 
container, as shown in Fig.\,\ref{fig:cross}. First, each of the modules is mounted on a trolley, which is itself
mechanically connected by a pivot link to a linear actuator (SKF - CAHB10~\cite{ref:SKF}). This actuator allows to 
move the module carriage on the rails by a few centimeters, which is needed to insert small radioactive 
sources between modules during calibration. These displacements are monitored to an accuracy 
of better than 5\,mm by mechanical position sensors mounted on the ground rail of the detector. As 
such a total of six calibration air gaps of 30 $\pm$ 5\,mm  can be created sequentially on both sides 
of each module.\\
 
The calibration robot that straddles the whole detector along its longitudinal axis is equipped with a holder 
for radioactive calibration sources as well as four capacitive sensors BCS (M18BBH1-PSC15H-EP02~\cite{ref:BCS}). 
Each module contains aluminium reference pins and stainless steel screws located on its top. Three 
capacitive sensors allow to monitor the longitudinal position of the robot by detecting the 
module reference pins. The fourth capacitive sensor ensures that the air gap 
is sufficiently large by measuring the distance between the stainless steel screws. Once the calibration robot 
is positioned between two modules, the source holder can further be moved along the X- and Y-axes. 
As such it can scan an area of 6 cells on the left and right sides of the plane center and 6 and 4 cells 
respectively above and below the plane center, covering nearly half of the detection plane's surface (see Fig.\,\ref{fig:cross}). 
The radioactive source (see Tables.~\ref{tab:neutronsource} and ~\ref{tab:gammasource}) are installed manually on the calibration arm from the outside of the container and the shielding, and are removed from the detector during normal data taking.

\section{Data Acquisition system}\label{sec:daq}

\subsection{Readout system design}\label{sec:readout_design}

The readout system is custom-made and based on a combination of analogue/digital front-end electronics and Field-Programmable Gate Array chips (FPGA). 
It brings together compactness, low power consumption (< 1\,kW), flexibility and high reliability for unattended operation on restricted access. All MPPC signals are equalized, synchronized (< 1\,ns) and continuously 
digitized at 40\,Msample/s. The use of zero suppression techniques (ZS), combined with pulse shape
 trigger algorithms, results in a data reduction factor of around 10\,k, down to 20\,Mb/s, with negligible dead time 
 (see Tab. \ref{tab:TriggerRates}).\\

\begin{figure}[h!]
\centering
\includegraphics[width=1.\textwidth]{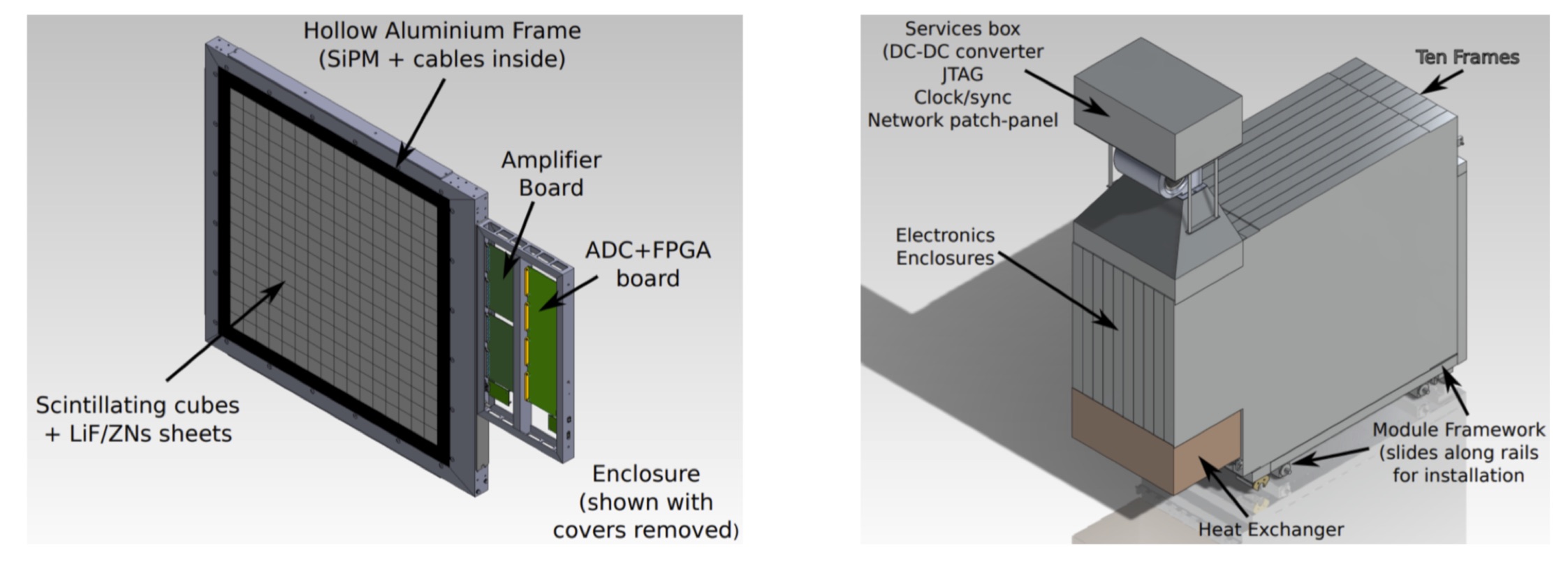}
\caption{{\small{ (Left) CAD rendering of a detector plane and its aluminium electronics enclosure. 
The 64 MPPCs are connected via an interface place using twisted-pair ribbon cables that terminate 
into insulation displacement connectors. (Right) Diagram of a ten planes detector module with its services box and 
its heat exchanger, placed below to take off heat generated by the electronics~\cite{Abreu:2018njy}.}}}
\label{fig:Daq_meca_design}
\end{figure}

The readout system operates on three levels: plane, module 
and full detector. Each of the 50 single detection planes has its own readout system, mounted 
directly on its side within a dedicated aluminium enclosure (see Fig.\,\ref{fig:Daq_meca_design}). 
It contains all the front-end electronics to run in autonomous mode, as described below. Each detector module is equipped with a heat exchanger and a services box that contains a DC-DC voltage converter to power the module, clock and synchronization distribution board, network patch panel and Minnow JTAG programming system. The module clock board (master/slave mode) provides a common clock fan-out to synchronise the ten associated 
digital boards. A master clock-board allows to run  the five detector modules synchronously.\\

The front-end electronics of a single detection plane consists of two 32-channel analogue boards, 
a 64-channel digital board, together with a power distribution system and an Inter-Integrated Circuit module 
that reads out four environmental sensors mounted inside the hollow frame. These environmental sensors monitor 
temperature and humidity levels throughout the detector. The two analogue boards are connected
 to the cathodes of the 64 MPPCs of the plane. They provide a common 70\,V power supply, as well as individual 
 trim bias voltages~(0-4\,V) used to equalize the amplitude response of each MPPC individually (see section\,\ref{sec:equalisation}). 
 Before being sent to the digital boards and in order to perform more accurate time stamp and amplitude measurements, 
the fast MPPC pulses (a few ns) are read out in differential AC coupled mode, amplified, band-pass filtered and shaped by 
a charge integrating operational amplifier to stretch the signal over several digital samples of 25\,ns each.\\

The two analogue boards are connected to a 64-channel digital board for digitisation and trigger. 
Each digital board has eight 8-channel ADCs, operating at a rate of 40\,MHz 
with 14\,bit resolution. Digital boards are controlled and read out over a 1\,Gbit/s optical Ethernet connection. 
A Phase-Locked Loop is included, which allows the digital  boards  to operate in standalone  mode  
using  an internally  generated  clock,  or  run  synchronised  to  an external  clock  signal. 
Triggers and readout logic are implemented in a Xilinx Artix-7 (XC7A200) 
based FPGA device~\cite{ref:XILINX}. JTAG connectors are included for remote firmware programming. Trigger signals 
from each digital board are propagated to all other detector planes by using two duplex 2.5\,Gbit links 
(copper cables). A complete description of the detector electronics is given in~\cite{Abreu:2018njy}.\\

The entire readout electronics is coupled very close to the detector, within aluminium 
enclosures, inside the chilled container. Both act as a Faraday cage, providing shielding 
from outside electronics noise. The top and bottom sides of these enclosures have openings 
to allow air flow cooling. The electronics are cooled by six fans mounted between the  services  
box  and  the  plane  electronics enclosures, pushing air downwards towards a heat exchanger 
which is capable of removing the 200\,W of heat generated by each module (see Fig.\,\ref{fig:Daq_meca_design}). 
The radiator unit is based on circulating water containing 18\% propylene glycol, connected 
to a chiller that operates nominally at a temperature of 5 degrees Celsius. It also acts as an overall 
cooling source to lower the ambient temperature inside the insulated detector container. As 
the environment temperature inside the container is maintained to 11 degrees Celsius, MPPC 
responses are stabilized at 1.4\% level and the MPPC dark count rate is reduced by a factor 
of three compared to operation at room temperature.

 \subsection{Online triggers and data reduction}\label{sec:trigger}

\begin{figure}[h!]
\centering
\includegraphics[width=.87\textwidth]{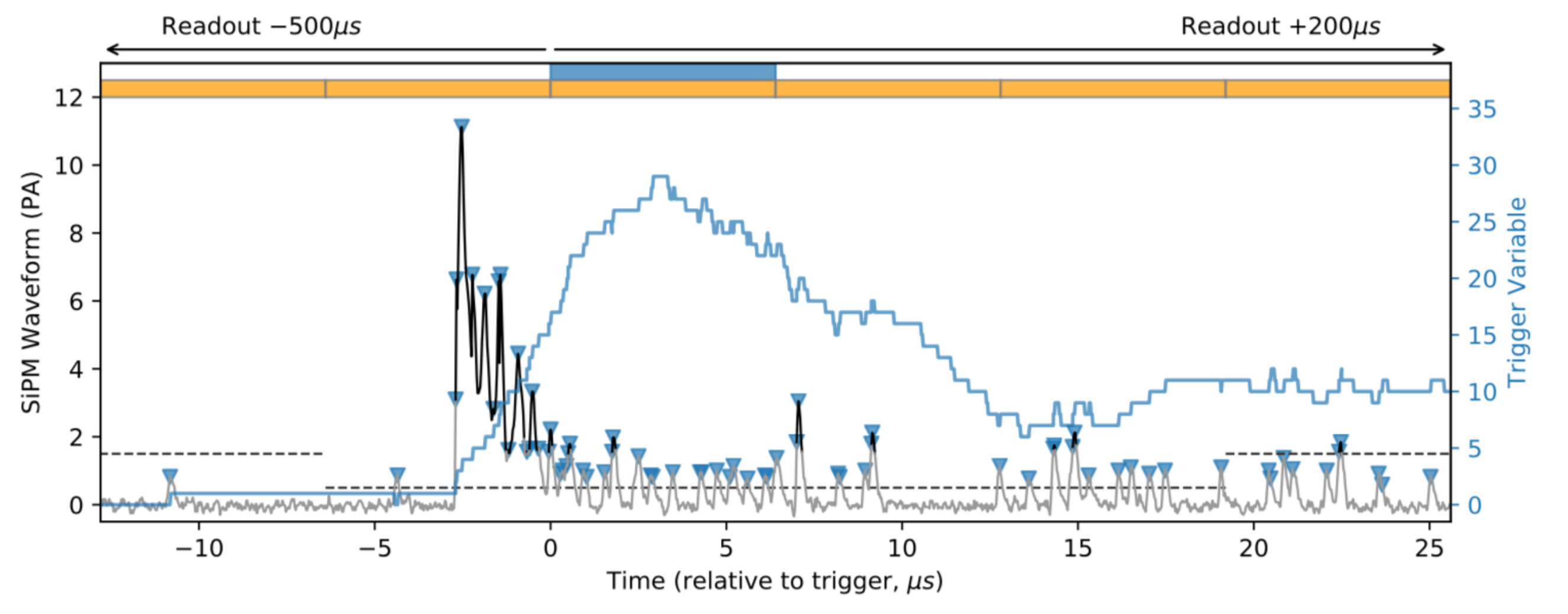}
\caption{{\small{Example of a {\em NS} waveform (black). The dashed lines show the 
zero suppression threshold. The value of the {\em NS} trigger variable, i.e number of peaks in the rolling 
time window, is shown in blue~\cite{Abreu:2018njy}.}}}
\label{fig:Neutron_wf}
\end{figure}

Multiple triggers and data reduction techniques have been implemented at the FPGA level \cite{bib:Arnold_2017}. The trigger strategy for neutrinos relies solely on triggering 
on a scintillation signal generated in the neutron detection screens, further denotes as {\em NS}. As the {\em NS} scintillation process is characterized 
 by a set of sporadic pulses emitted over several microseconds (see~section~\ref{sec:layout}), 
 the {\em NS} trigger algorithm involves tracking the time density of peaks in the waveform~\cite{Abreu:2018njy}. All algorithm parameters have been optimized during deployment: the 
 amplitude threshold on waveform local maxima to be counted as a peak is set to 0.5\,PA, the 
 size of the rolling time window is fixed at 256 waveform samples (6.4\,\textmu{}s) and the number 
 of peaks, required in the window, is set to 17 (see Fig.\,\ref{fig:Neutron_wf}). These default 
 values correspond to a trigger efficiency of 75\% and a purity of 20\%  during nominal reactor ON periods.
 The efficiency is defined by the ratio of triggered neutrons and the total number of captured neutrons, as determined from the calibrated activity of our calibration source source and a capture efficiency obtained from simulation. 
 The purity is defined as the number  of triggers passing an offline neutron identification and the total number of triggers. The offline neutron selection is demonstrated in Fig~\ref{fig:NS_calib1}, and it has a purity of 99\%. The 80\% non-neutron triggers are mostly muon signals, which can be distinguished using an offline identification (see section \ref{seq:SDQM} and \ref{sec:calibration}). For each {\em NS} trigger, a large space-time region is read out in order to encapsulate all signals from the IBD interaction. Three planes are read out on either side of the triggered plane, with a large time window of 500\,\textmu{}s before the trigger and 200\,\textmu{}s after the trigger. The {\em NS} trigger rate, which does not change significantly depending on reactor operation, fluctuates around 80\,Hz corresponding to a data-rate of 15\,MB/s (see Tab.~\ref{tab:TriggerRates}).\\

Two additional triggers are also implemented to measure background and to survey the detector stability. 
A threshold trigger has been implemented to record high amplitude {\em ES} signals, such as muons. The default 
physics mode threshold is 2\,MeV with a X-Y coincidence imposed. This gives a trigger rate of about 2.1\,kHz 
and data-rate of 2\,MB/s during nominal reactor ON periods. It decreases by around 10\% during reactor OFF periods (see Tab.\ref{tab:TriggerRates} and Fig.\,\ref{fig:triggers_trend}). A periodic trigger has also been implemented in order to monitor continuously the stability of the MPPCs, as well as any noise contributions. The entire detector is read out for a time window of 512 samples without zero suppression, with a default trigger rate of 1.2\,Hz, giving a data rate of 3.9\,MB/s (see Tab.\,\ref{tab:TriggerRates}). The three triggers include storing MPPC waveforms for offline analysis. A zero suppression value at 1.5\,PA, respectively 0.5\,PA in {\em NS} mode, allows to remove the pedestal contribution, whilst retaining all MPPC signals. It results in a waveform compression factor of around 50 (resp. 500) \cite{Abreu:2018njy}. Table~\ref{tab:TriggerRates} summarizes the different trigger parameters and their associated data rates.

\begin{table}[!h]
\centering
{\scriptsize{
\begin{tabular}{|c|c|c|c|c|c|c|} 
      \hline 
   Trigger  &      ZS       &       Condition     &  \multicolumn{2}{c|}{Readout Region}  & Trigger rate &  Data rate \\     \cline{4-5} 
     Type    & Threshold  &                          &  Space &  Time  ($\mu$s)                   &        (Hz)      &    (MB/s)  \\    
       \hline     \hline 
    Periodic &  Disabled   & Random~1.2\,Hz                          & Whole detector   &  12.8              & 1.2& 3.9 (19\%)\\      \hline 
   Threshold   & 1.5\,PA   & Waveform sample > 50\,PA     & Triggered plane  &   6.4                & 2100 & 2 (10\%)\\    \hline 
     {\em NS}           & 0.5\,PA    & N\textsubscript{peak}  $\ge $ 17 peaks          & Triggered plane   &   [-500,+200]  &  80  & 15  (71\%)\\     
                     &               & (  Width = 6.4\,\textmu{}s , T\textsubscript{peak} = 0.5\,PA )          & $\pm$ 3 planes  &                        &   &\\     \hline   
\end{tabular}}}
\caption{{\small  Summary of trigger settings and associated data rates during reactor ON physics data taking \cite{Abreu:2018njy}.}}
\label{tab:TriggerRates}
\end{table}

The readout software runs on a disk server, located very close to the detector. It provides 50\,TB of local storage, that 
is split into two data partitions, which are periodically swapped and cleared. All the data are first transferred to the 
Brussels HEP Tier 2 data centre, then subsequently backed up at CC-IN2P3 in France \cite{bib:ccin2p3} and at 
Imperial College in the UK using GRID tools, which are used for offline processing and simulation production.

\section{The BR2 reactor at SCK$\cdot$CEN}\label{sec:BR2}

\subsection{The BR2 reactor}\label{sec:BR2_description}
 
\begin{figure}[h!]
\centering
\includegraphics[width=.99\textwidth]{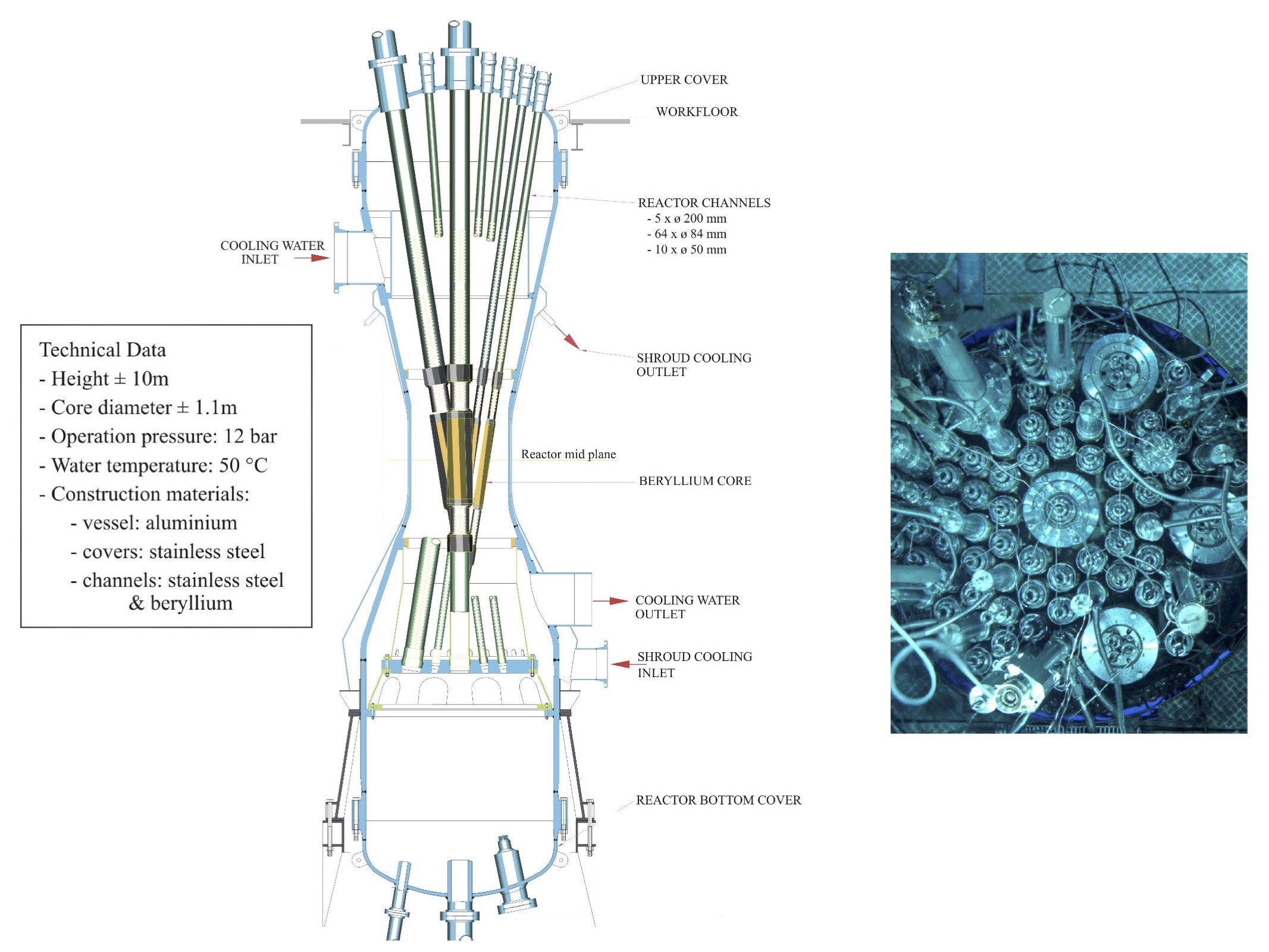}
\caption{{\small{(Left) Design and technical data of the BR2 reactor core. It consists of a 
beryllium matrix composed of 79 hexagonal channels containing the nuclear fuel 
elements, the control rods and the experimental channels. The beryllium core (yellow) 
is confined in an aluminium vessel (blue), that is completely under water. (Right) Picture of the 
upper cover of the reactor vessel \cite{ref:BR2_web}.}
}}
\label{fig:BR2_sketch}
\end{figure}

The BR2 reactor (Belgian Reactor 2) is a materials testing reactor operated by the nuclear research center SCK\raisebox{-0.8ex}{\scalebox{2.8}{$\cdot$}}CEN in Mol (Belgium). Since its start-up in 1963, it is one of the most powerful research reactors in the world and thus plays an important role in nuclear material and fuel R\&D. It is also widely used for production of medical isotopes and neutron transmutation doped silicon~\cite{ref:BR2_web}. The BR2 reactor is a pressurized "tank-in-pool" type reactor, cooled with water and moderated by its beryllium structure and water (see Fig.~\ref{fig:BR2_sketch}). It has a unique twisted design with inclined channels to obtain a compact core. The BR2 reactor uses highly enriched uranium fuel (HEU: 93.5\% $^{235}$U) at powers varying between 40 and 100\,MW$_{th}$. It thus produces a very high neutron flux, up to 10$^{15}$ n/cm$^{2}$/s, and provides an intense source of antineutrinos up to about 2$\cdot$10$^{19}$\,$\bar\nu_e/$s.\\

At the end of the SM1 prototype physics run, the BR2 reactor was shut down for a 
period of one year and a half, and has undergone a thorough overhaul. The BR2 
operation was restarted in July 2016. In practice, the reactor operates at a nominal 
power of about 65 MW$_{th}$, for 160 to 210 days per year, 
during cycles of about three to four weeks (ON period). There are on average 6 cycles of reactor 
ON periods per year, that alternate with interim maintenance periods of the same duration (OFF period). 
The Solid experiment takes advantage of the OFF periods to perform calibration campaigns 
and background measurements (see section \ref{sec:calibration}).. 

%
 \subsection{Detector integration on site}\label{sec:integration}
%

\begin{figure}[h!]
	\centering
	\includegraphics[width=.9\textwidth]{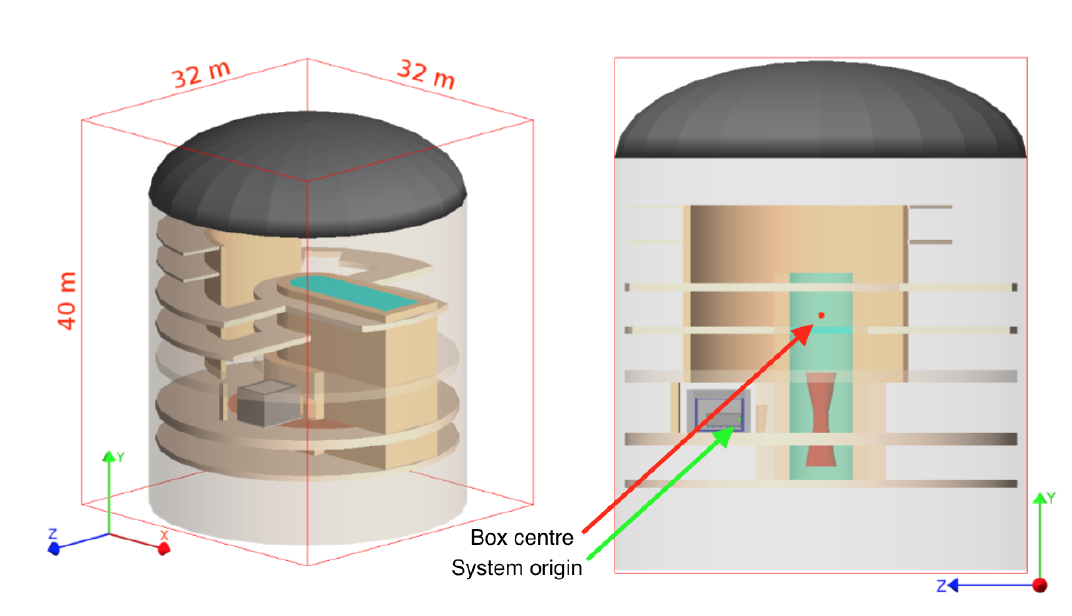}
	\caption{{\small{3D representation of the BR2 geometry model and positioning of the SoLid 
	detector as implemented in SoLidSim (see section\,\ref{sec:simulation}). The SoLid position system 
	is based on three Cartesian coordinates along perpendicular axes in a right-handed system. The Z-axis 
	is perpendicular to the detector planes and its direction points away from the nominal center of the 
	BR2 reactor core. The Y-axis points upward towards the zenith, and the X-axis points to the right side 
	of the detector, when facing the reactor.}}}
	\label{fig:br2sim}
\end{figure}

\begin{figure}[h!]
	\centering
	\includegraphics[width=.75\textwidth]{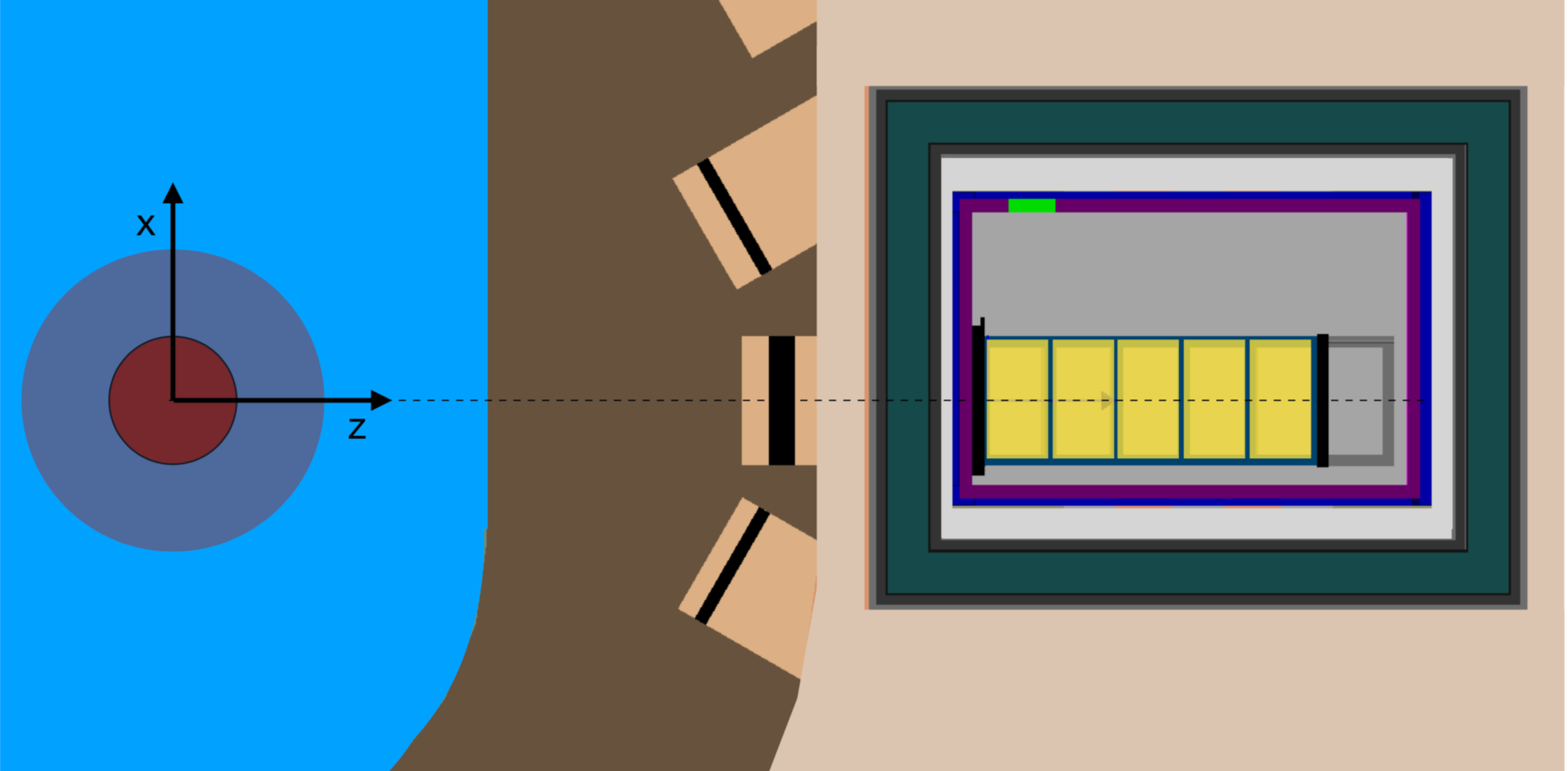}
	\caption{{\small{A vertical projection of the detector geometry and its positioning in the containment 
	building (\GEANTfour based). It shows the reactor core (red) submerged  in water (blue) and the detector 
	geometry including the detector module placement (yellow and blue rectangles) inside the cargo container, 
	the rail system (dark grey rectangle), container insulation (purple) and passive water shielding (dark green). }}}
	 \label{fig:phase1_projected}
\end{figure}

The SoLid detector is located at level 3 of the BR2 containment building in direct line-of-sight of the nominal reactor core center. This is the third detector installed at this location by the collaboration, after the two prototypes, NEMENIX \cite{Abreu:2017bpe} and SM1\cite{Abreu:2018pxg}. The 50 detector planes are oriented perpendicularly to the detector-reactor axis, and as close as possible to the reactor core. As such, the sensitive volume of the SoLid  detector covers a baseline of $[6300\,\text{mm} - 8938~\text{mm}]$ away from the nominal center of the BR2 reactor core (see Fig.\,\ref{fig:br2sim} and Fig.\,\ref{fig:phase1_projected}). As the aluminium reactor vessel is totally immersed in water, its radiation is properly shielded. Moreover, at this floor of the containment building, no other experiments surround the detector and all neighbouring beam ports have been shielded with 20 cm thickness of lead. It thus ensures stable and low reactor induced background conditions.\\

\begin{figure}[h!]
	\centering
	\includegraphics[width=1.\textwidth]{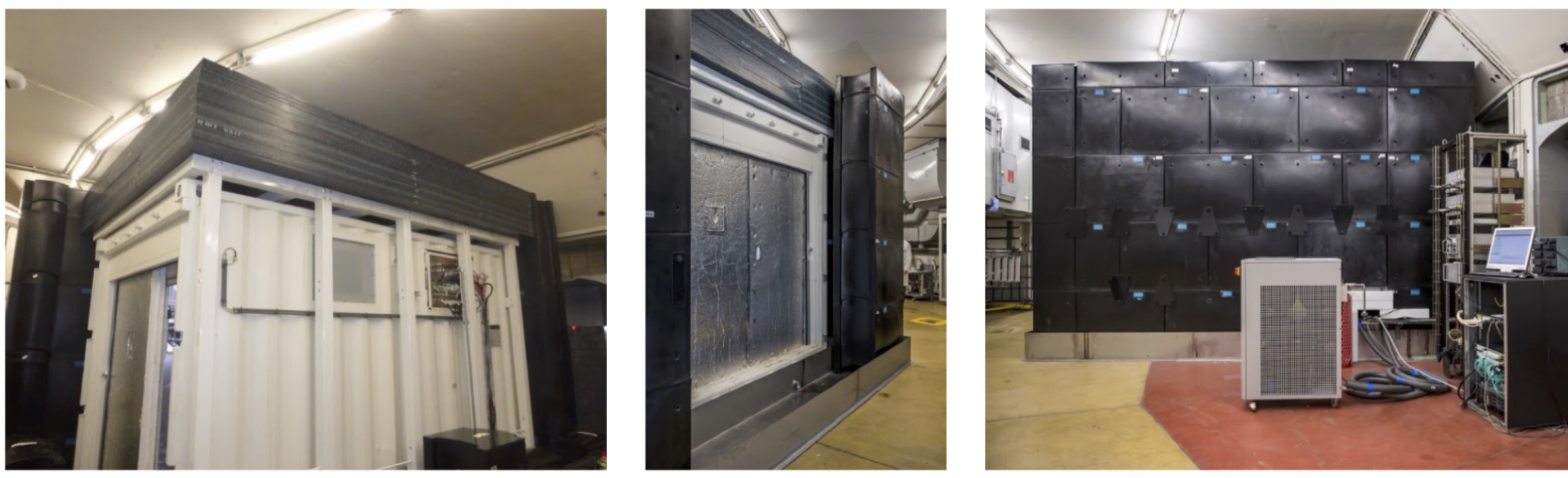}
	\caption{{\small{Pictures of the detector during its integration at BR2: cooled container (white) and passive shielding (black). The power supplies and DAQ system, together with the chiller used for the container cooling are visible on the side of the detector (front of picture).}}}
	\label{fig:integration}
\end{figure}

The overburden above the detector is composed of 3 concrete floors and the steel roof of the containment building  (see Fig.\,\ref{fig:br2sim}). It corresponds to 8 meters-water-equivalent. In order to mitigate the atmospheric and cosmic backgrounds, which were determined experimentally with SM1 and compared with a full-chain \GEANTfour-based Monte-Carlo simulation~\cite{ref:Pinera-Hernandez:2016qkg}, a passive shielding surrounds the detector (see~Fig.\,\ref{fig:phase1_projected}~and~Fig.\,\ref{fig:integration}). It is maximized for cost, avalaible space and floor load versus attenuation of cosmic neutrons. The top of the detector is shielded with a 50\,cm PE layer made of 2.5\,cm thick PE slabs that are staggered to avoid gaps. The PE slabs are supported by a steel scaffolding straddling the container and surrounded by a 50\,cm thick water wall on the four sides of the container. The cosmic neutron flux in the energy range [1-20\,MeV] is thus reduced by a factor of 10 and about 5\% are converted to slow neutrons (E$_n$<10 eV) that penetrate the wall. In order to capture these slow neutrons, thin cadmium sheets with a thickness of 2\,mm are sandwiched between the passive shielding and the container housing of the detector. The capture efficiency of these cadmium sheets for slow neutron is about 88\%. The cadmium sheets, cover the entire back side of the experiment container and most of its top and bottom surface, amounting to roughly 45\% coverage of the experiment.\\

The environment of the BR2 containment building is continuously monitored and registered by the 
BR2 Integrated Data Acquisition System for Survey and Experiments (BIDASSE). 
During  SoLid operation, environmental parameters, such as temperature, humidity and pressure, 
outside and inside the containment building, are constantly monitored. Also the background radiation 
is monitored using gamma and beta detectors placed in the vicinity of the SoLid container. So far, 
these variables are used as a cross check of the data 
coming from the container instrumentation, i.e. environmental sensors, NaI scintillator and airborne 
radon detector mentioned in section\,\ref{sec:container}.\\

The thermal power is determined by measuring the flow rates and temperatures at the entrance and inlet of the primary cooling circuit and the reactor pool circuit. The flow of the cooling water is measured using Dall tubes and the temperature is measured using resistor thermometers. The main uncertainty in this measurement originates from the calibration of the Dall tubes.
The difference between the Dall flow measurement and the measurements performed by the I.A.E.A. using ultrasonic equipment at another location in the primary circuit is equal to 4.8\%. The signal processing chain introduces some possible systematic offsets as well, which results in a conservative uncertainty estimate of 5\% on the thermal power determination.\\
%
%

 \subsection{Neutrino flux modeling}\label{sec:BR2_nu_flux}

\begin{figure}[h!]
	\centering
	\includegraphics[width=.49\textwidth]{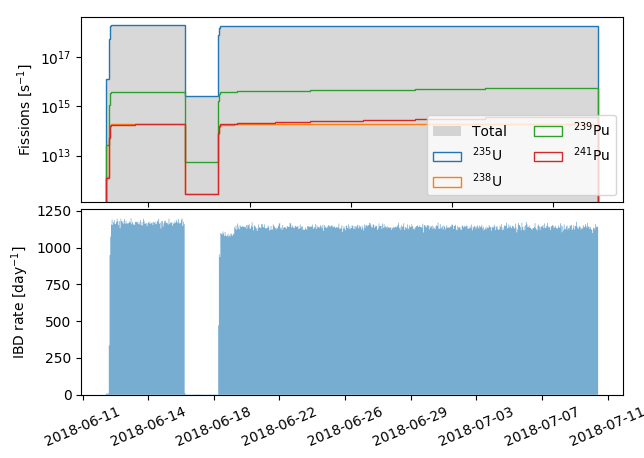}
	\includegraphics[width=.49\textwidth]{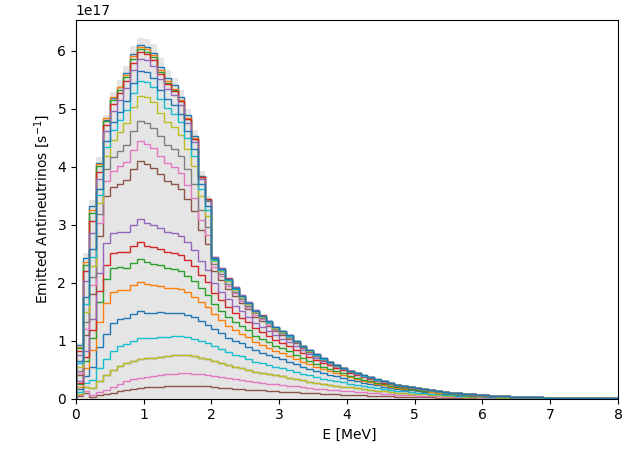}
	\caption{{\small{(Left) Evolution of the fission rates (top) and the associated IBD interaction rate (bottom) during cycle 03/2018A. (Right) Emitted antineutrino spectrum calculated with MCNPX/CINDER90 fission rates and using the summation method. The different colors correspond to different time steps before reaching equilibrium, from 0 to 28 days of irradiation time.\cite{Silvapaper}.}}}
	\label{fig:corecomputation}
\end{figure}

For each cycle, i.e. for a given fuel loading map and operation history, detailed simulations of the BR2 reactor core are performed to calculate the emitted antineutrino spectrum. In addition, the computation of the spatial fission distribution, combined with a dedicated tracking algorithm, allows to obtain the detector acceptance, defined as the fraction of emitted antineutrinos that
pass through the detector. The RMS of the neutrino emission point distribution within the reactor core are about 50 \,cm in diameter and 80 \,cm in height. The geometrical detector acceptance is about 0.11\%. It depends slightly on the fuel loading map. The emitted antineutrino spectrum is computed using the conversion \cite{PhysRevC.83.054615,Mention_2011} and summation methods \cite{Silvapaper}. The conversion method is based on the prediction of the fission rates as a function of time using a MCNPX \cite{ref:MCNPX} (or MCNP6\cite{ref:MCNP6}) 3D model of the reactor core interfaced with the evolution code MCNPX/CINDER90 \cite{ref:MCNPX} and combined with the converted $\beta^{-}$ spectra, measured at ILL reactor in Grenoble, France (see Fig.\,\ref{fig:corecomputation}). However, since ON/OFF reactor transitions are frequent and reactor ON cycles are relatively short, we have to take into account off-equilibrium effects. To do so, the MURE code \cite{ref:MURE} allows to adapt the converted spectra to the irradiation time of the antineutrino experiment. The summation method uses the same MCNPX/CINDER90 software combined with the amount of in-core $\beta^{-}$ emitters and consists in summing all the individual beta branches composing the total spectrum weighted by the beta decay activities \cite{ref:FALLOT} (see Fig.\,\ref{fig:corecomputation}). Systematic effects coming from the thermal power uncertainty, modeling uncertainty as well as nuclear data, will also be estimated. The current calculations indicate that at typical power settings of BR2, the SoLid experiment is subjected to IBD interaction rates between 11.5 and 14.5\,mHz (1000-1250 detectable events per day).

%

\subsection{Backgrounds}\label{sec:BR2_bg}

The SoLid detector is also subjected to various background processes that contaminate the IBD samples for final analysis. Because the primary physics trigger is set to detect thermal neutrons interacting in the neutron detection screens, most backgrounds are related to either the production of neutrons via processes other than IBD interactions, or processes that excite the ZnS(Ag) scintillator embedded in the neutron detection screens. Some background processes exhibit a clear time structure between the triggered {\em NS} time and preceding {\em ES} signals and are called {\em correlated}. Others have a random time structure and are called {\em accidental}. Reactor independent backgrounds dominate our data sample and can be extracted from data collected during reactor OFF periods. We quantify our understanding of these background components by comparing background simulations with data in specific control regions that are enriched in one specific background component. Reactor dependent backgrounds are very scarce and are monitored using a dedicated NaI gamma ray detector and with dedicated control samples that are depleted of IBD events. It is mostly composed of gammas, and thus only populate accidental events, i.e. a random coincidence of a {\em NS} and {\em ES} signal within the IBD trigger window.
In all cases we try to validate the background composition and the influence of selection criteria by using dedicated Monte Carlo simulations, wherever they are available. A detailed description and treatment of these models falls beyond the scope of this paper and will be described at length in a following physics analysis paper. Here we summarize the main background processes and their origin. \\

A first source of neutrons to which the detector is constantly exposed is of atmospheric origin. These neutrons are produced by cosmic ray spallation when high energy primaries collide with atmospheric nuclei. Neutrons can penetrate much further into our atmosphere than the electromagnetic component and are shown to produce a complex energy spectrum~\cite{gordon} ranging from sub-eV to multi-GeV. The flux of atmospheric neutrons is simulated using the Gordon model as described in~\cite{gordon}, scaled to the BR2 reactor site elevation and latitude, and cross-checked with the more general purpose CRY generator~\cite{CRY}. The flux contains slow and fast neutrons that induce a different response in the SoLid detector.
Slow neutrons that enter our detector can, in combination with an accidental coincidence of an {\em ES} signal such as those induced by gamma rays, produce signals similar to IBD events. The detector timing and spatial segmentation with corresponding topological selections can largely suppress this background. The passive water shield of 50\,cm surrounding the experiment, combined with the Cd sheets placed on the outer walls of the container help to thermalize and capture some of the epithermal neutrons.
The fast neutron component is able to penetrate the detector and can induce highly energetic proton recoils resulting in {\em ES} signals. If the neutron further thermalizes inside the detector it can be captured and induce a {\em NS} trigger. As such it introduces a time correlated background that dominates the selected IBD events samples for  {\em ES} signals with energy above 5\,MeV. This background is mainly suppressed by timing and {\em ES} signal multiplicity requirements.\\

Cosmic ray muons are also known to induce spallation reactions in materials near or inside the SoLid detector that produce neutrons or radioisotopes. The rate of neutron production increases with muon energy and with material density. The rate and spectrum is modelled using the CRY generator~\cite{CRY} by simulating cosmic ray showers on a surface that lies 30\,m above the BR2 building and by tracking all shower components through the building and detector geometry. Roughly one third of the spallation neutrons are produced inside the detector, while the rest is created in surrounding structures. The techniques to mitigate the corresponding accidental and time correlated background are similar to those to reduce the atmospheric neutron background. Cosmic muons themselves are used as a calibration tool, as they generally leave a reconstructed track in the detector. In some cases, however, muons can clip the detector edges, leaving an isolated energy deposit that can contribute to the accidental backgrounds in the detector. Muons can also decay in the detector, resulting in the detection of the Michel electron or positron with a characteristic delay corresponding to the muon life time. The rate and spectrum of cosmic ray muons are modelled using CRY, but are cross-checked by other models by Guan~\cite{Guan:2015vja} and Reyna~\cite{arXiv:hep-ph/0604145}.\\

Intrinsic radioactivity of detector materials or airborne isotopes are another source of backgrounds. The 
airborne isotope of $^{222}$Rn can produce several alpha and beta particles along its decay chain. 
Its presence inside the detector container is therefore monitored by a dedicated Rn detector, as described in section~\ref{sec:container}.  Another source 
of intrinsic radioactivity are trace fractions of Bi isotopes contained in detector materials, in particular the 
neutron detection screens. The $^{214}$Bi isotope is the most troublesome and is part of the long $^{238}$U 
decay chain. It decays to $^{214}$Po via $\beta^{-}$ emission with a half-life of roughly 20 minutes and a Q$_{\beta}$ value of 3 MeV. The resulting Po isotope has a half life of 164~\textmu{}s and emits an energetic alpha 
particle that can cause a scintillation of the ZnS(Ag) scintillator of the neutron detection screens. The half life of $^{214}$Po 
is very similar to the thermalization and capture time of fast neutrons in the SoLid detector. This background, 
referred to as BiPo, dominates at prompt energies below 3 MeV and is difficult to mitigate. This BiPo 
background is modelled by generating random decay vertices in the neutron detection screens throughout 
the detector, followed by the subsequent decays with corresponding half lifes and energies. 
The use of cube and fibre topology information allows to localize the spatial origin of the alpha particle, 
while timing and energy can be used to tag the {\em ES} signal. In addition, also the integrated energy of the 
{\em NS} signal can be used to discriminate neutrons and alphas from the $^{214}$Po decay.\\

Other backgrounds can be broadly categorized as accidentals and consist of random coincidences of {\em ES} signals that are typically induced by gamma rays and thermal neutrons in the surroundings of the detector.  The accidental distribution can vary with reactor power, but can be easily extracted from data itself, using negative time differences between the {\em ES} and the {\em NS} signals. Accidentals contribute only marginally to the selected 
IBD events sample.
 
\section{Detector operation and data monitoring}\label{sec:monitoring}

\subsection{Channel characterization and equalization}\label{sec:equalisation}

\begin{figure}[h!]
\centering
\includegraphics[width=.95\textwidth]{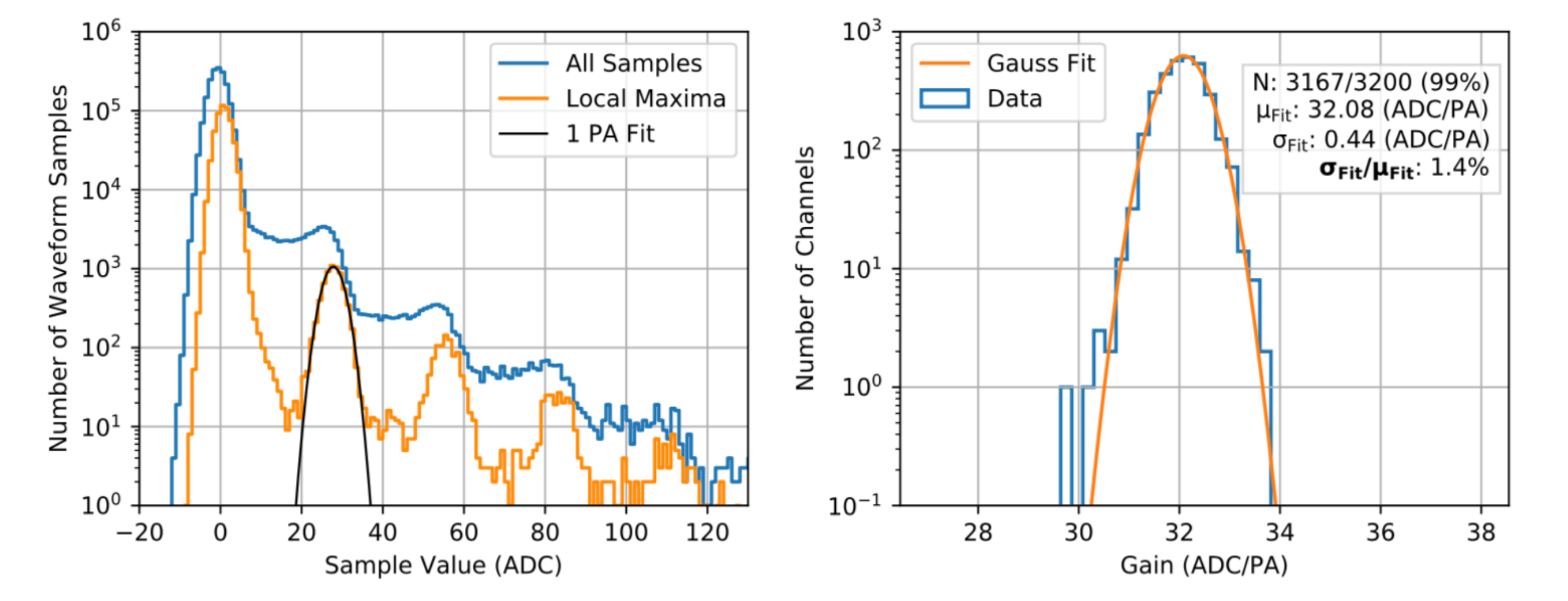}
\caption{{\small  (Left) Spectrum of ADC samples for a typical detector channel, with and without a local 
maxima filter applied. The first pixel avalanche peak can be parametrized using a Gaussian curve (shown in black), and 
the Gaussian mean value is used as the channel gain measurement. (Right) Spread of gain values across all 
operational MPPCs after the final iteration of the equalization procedure \cite{Abreu:2018njy}.}}
\label{fig:ADC_equalisation}
\end{figure}

During nominal operations, the average gain of the MPPCs is set to 32.1 ADC counts per PA, which corresponds to a mean 
over-voltage of 1.8\,V above the avalanche breakdown value of each sensor. This over-voltage setting 
was optimized for neutron efficiency during the commissioning of the detector at BR2. It is a compromise 
between photon detection efficiency, pixel cross talk and thermal dark count rate. The amplitude response 
of the sensors is equalized by an automatic procedure that first consist of finding the individual 
break down voltage of each MPPC, which is spread with a standard deviation of around 2\,V over all the sensors. 
For a given channel, the linear relationship between gain and voltage is determined by performing a voltage 
scan. This procedure allows to equalize the gain of all the channels with a spread around 1.4\%, where the 
dominant uncertainty is the precision of the gain-finder itself (see Fig.\,\ref{fig:ADC_equalisation}).\\

\begin{figure}[h!]
\centering
\includegraphics[width=1.\textwidth]{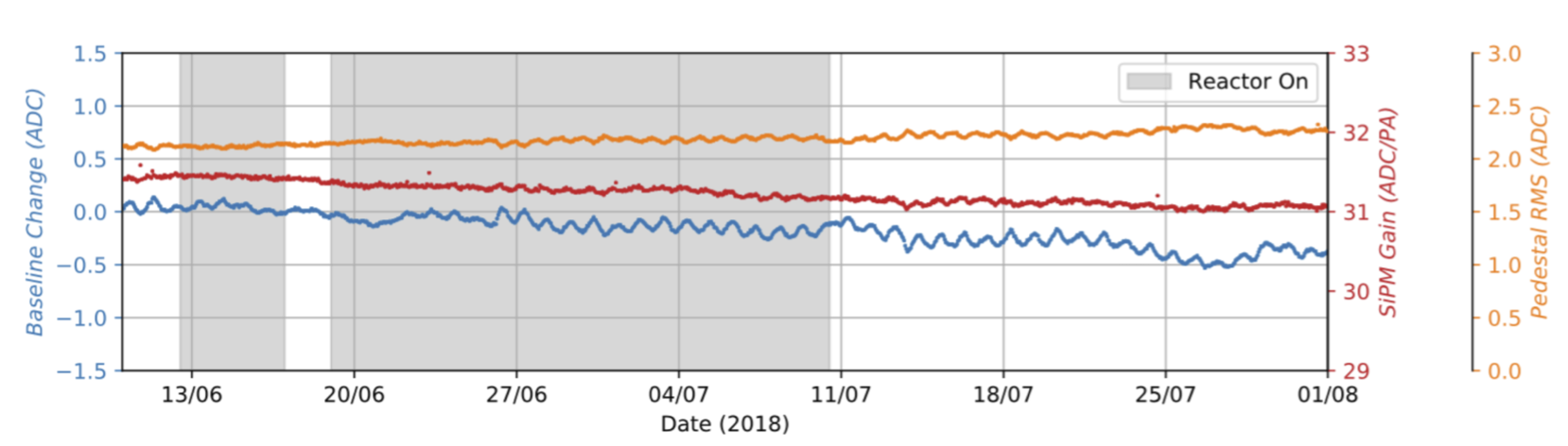}
\caption{{\small Weekly trends of the MPPCs response. Day-night variations are observed 
with temperature which changes of up to 0.5$^{\circ}$C inside the container. The small 
increase in average temperature over the period of data taking  increases the 
dark count rate, causing the baseline values to change by up to 2\% relative to the gain \cite{Abreu:2018njy}. Reactor cycles are indicated by grey bands and have no observable influence on the operational stability of the detector.}}
\label{fig:MPPC_response}
\end{figure}

MPPC sensors typically have a high dark count rate, which is the main reason why the detector 
is cooled inside an insulated container. The rate also strongly depends on the over-voltages applied. Under 
nominal running conditions, i.e. at a mean over-voltage of 1.8\,V and at a temperature of 11$^\circ$C, the 
mean dark count rate is 110\,kHz per channel, which is uniform across the detector. The MPPC pixel cross talk, 
which corresponds to the probability that a pixel avalanche triggers an avalanche in a neighbouring pixel, also 
depends on the bias voltage and amounts to 20\% for an over-voltage of 1.8\,V \cite{Abreu:2018njy}. Long 
term trends of the MPPC demonstrate a stable operation, as shown on Fig.\,\ref{fig:MPPC_response}. 

\subsection{Detector operation and data quality monitoring}\label{seq:SDQM}

\begin{figure}[h!]
	\centering
	\includegraphics[width=.99\textwidth]{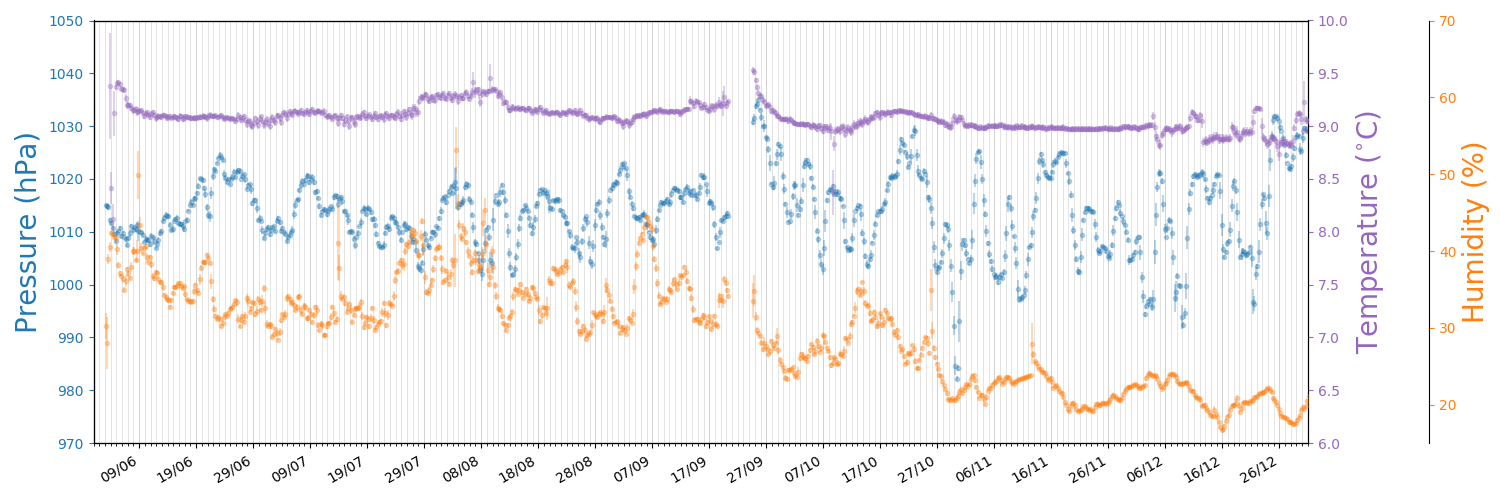}
	\caption{{\small Time evolution of some relevant environmental parameters, measured by sensors placed in or near the SoLid detector.}}
	\label{fig:env_trend}
\end{figure}

Run operations are controlled via a dedicated Python-driven web application, 
the "SoLid Data Quality Monitor" (SDQM). It automatically processes a small fraction of each run (first GB) 
using the SoLid reconstruction and analysis software. Output measurements 
and distributions of the detector as well as in-situ environmental sensors are read out periodically, 
as show in Fig.\,\ref{fig:env_trend} and stored in an online database and is continuously inspected via a web application.\\ 

\begin{figure}[h!]
\centering
\includegraphics[width=.9\textwidth]{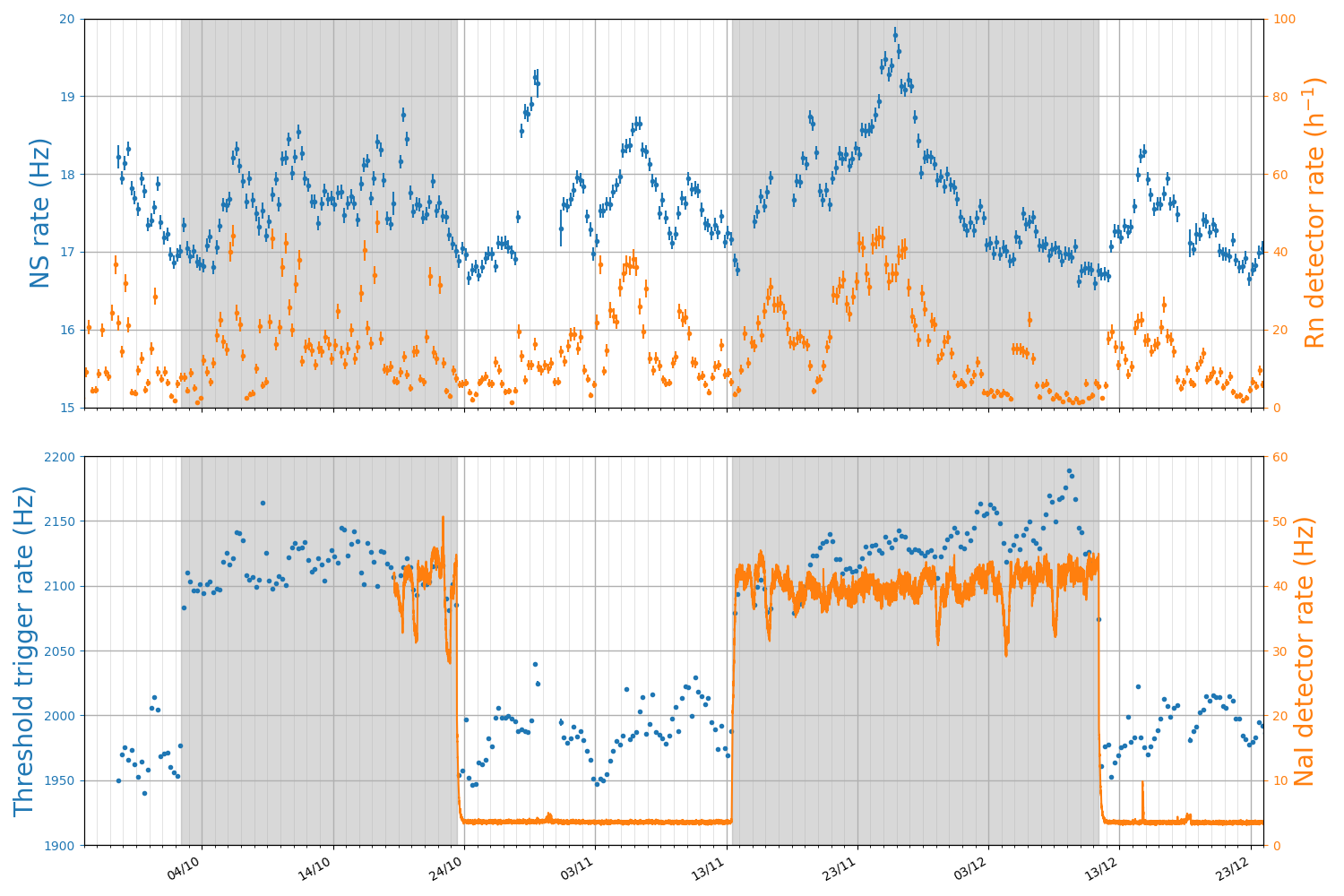}\vspace*{0.1cm}
\caption{{\small (Top) Long term trends of the {\em NS} rate after muon contamination removal (blue) and the airborne radon detector rate (orange). (Bottom) Long term trends of the threshold trigger rate (blue) 
and the NaI detector rate (orange). Reactor ON periods are shown as grey bands.}}
\label{fig:triggers_trend}
\end{figure}

The rates obtained from the monitoring database are shown in Fig.\,\ref{fig:triggers_trend}. 
The {\em NS} trigger rate stays stable irrespective to the reactor operation. Once the muon contamination 
has been removed, the {\em NS} rate is around 18 Hz and is strongly correlated to the airborne radon 
concentration which is monitored by a Rn detector (see section\,\ref{sec:container}). 
The transition between the reactor ON and OFF periods can only be seen by the relatively small change in the threshold trigger rate, which is strongly correlated to the gamma rate measured by the NaI detector placed inside the container next to the detector.~\\

\begin{figure}[h!]
\centering
\includegraphics[width=.9\textwidth]{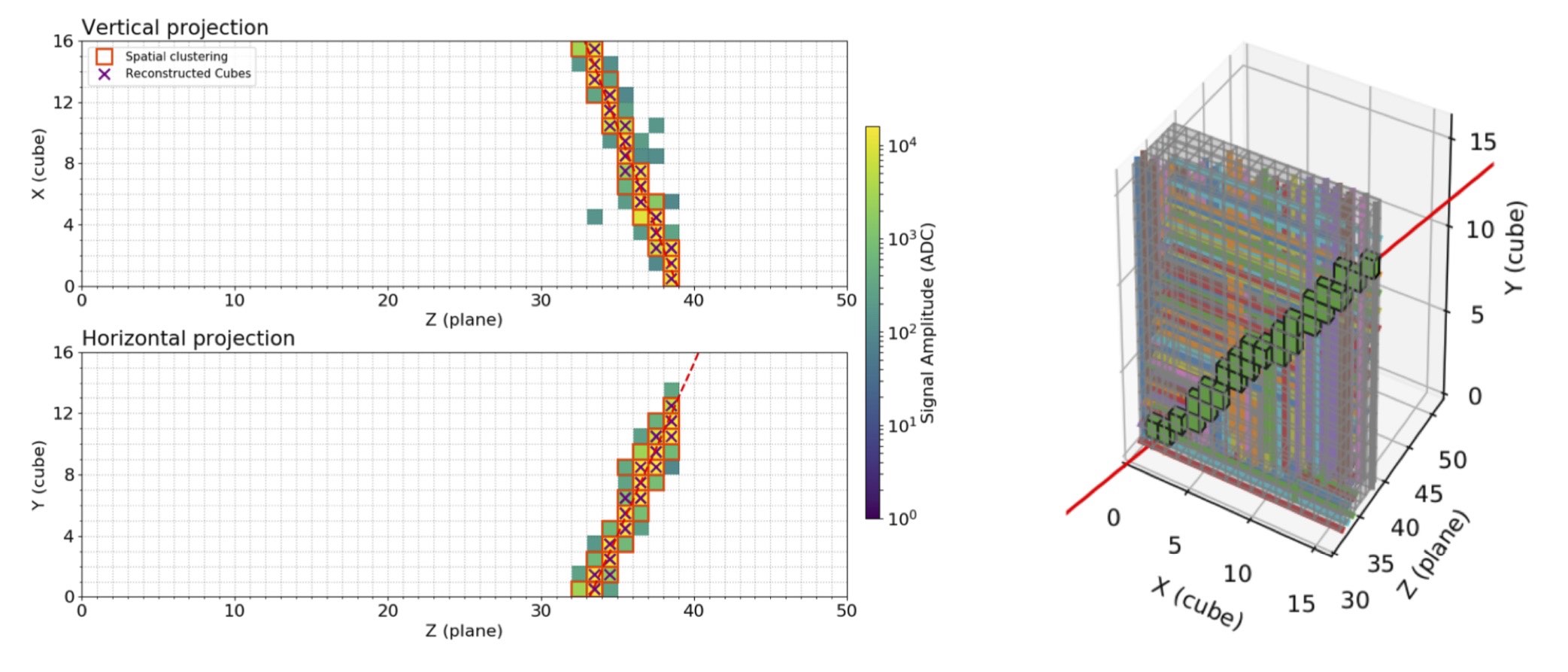}\vspace*{0.1cm}
\caption{{\small An example of a reconstructed muon track inside the detector (red lines). 
Red squares indicate which channels are used for the track reconstruction, the purple crosses
shows through which cells the reconstructed track goes.}}
\label{fig:muons_track}
\end{figure}

The SoLid detector segmentation provides a powerful tool for identifying cosmic muons crossing 
the detector. Muons deposit their energy in a large number of cells along their path. 
Their offline reconstruction thus relies on a spatial clustering that groups all signals from neighbouring 
fibres, an energy requirement to reject low energetic secondary signals, and finally, a requirement on the fibre 
multiplicity. An example of a reconstructed muon track inside the detector is displayed in Fig.\,\ref{fig:muons_track}.\\

\begin{figure}[h!]
\centering
\includegraphics[width=.99\textwidth]{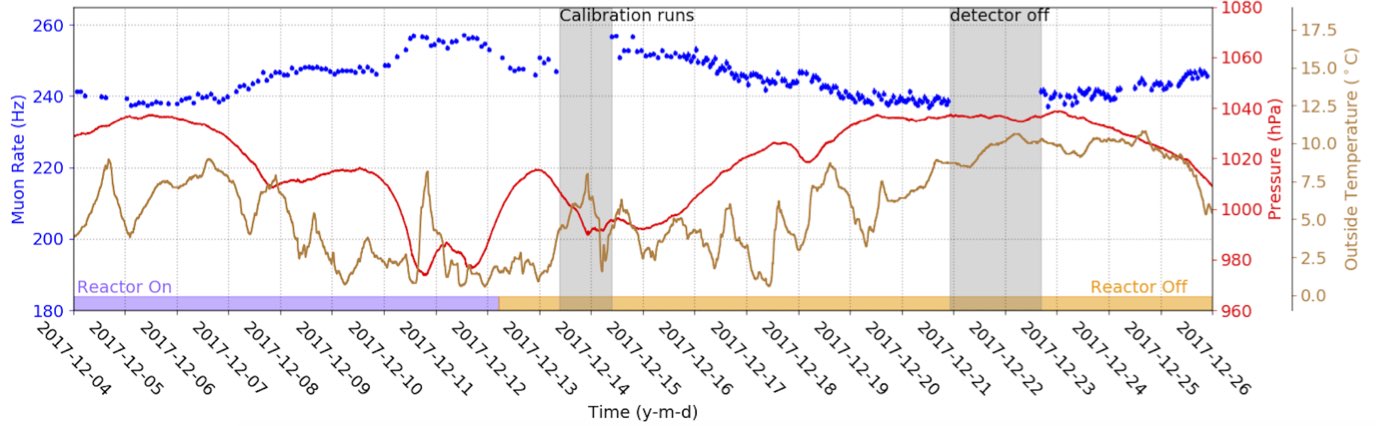}\\ \vspace*{0.1cm}
\includegraphics[width=.675\textwidth]{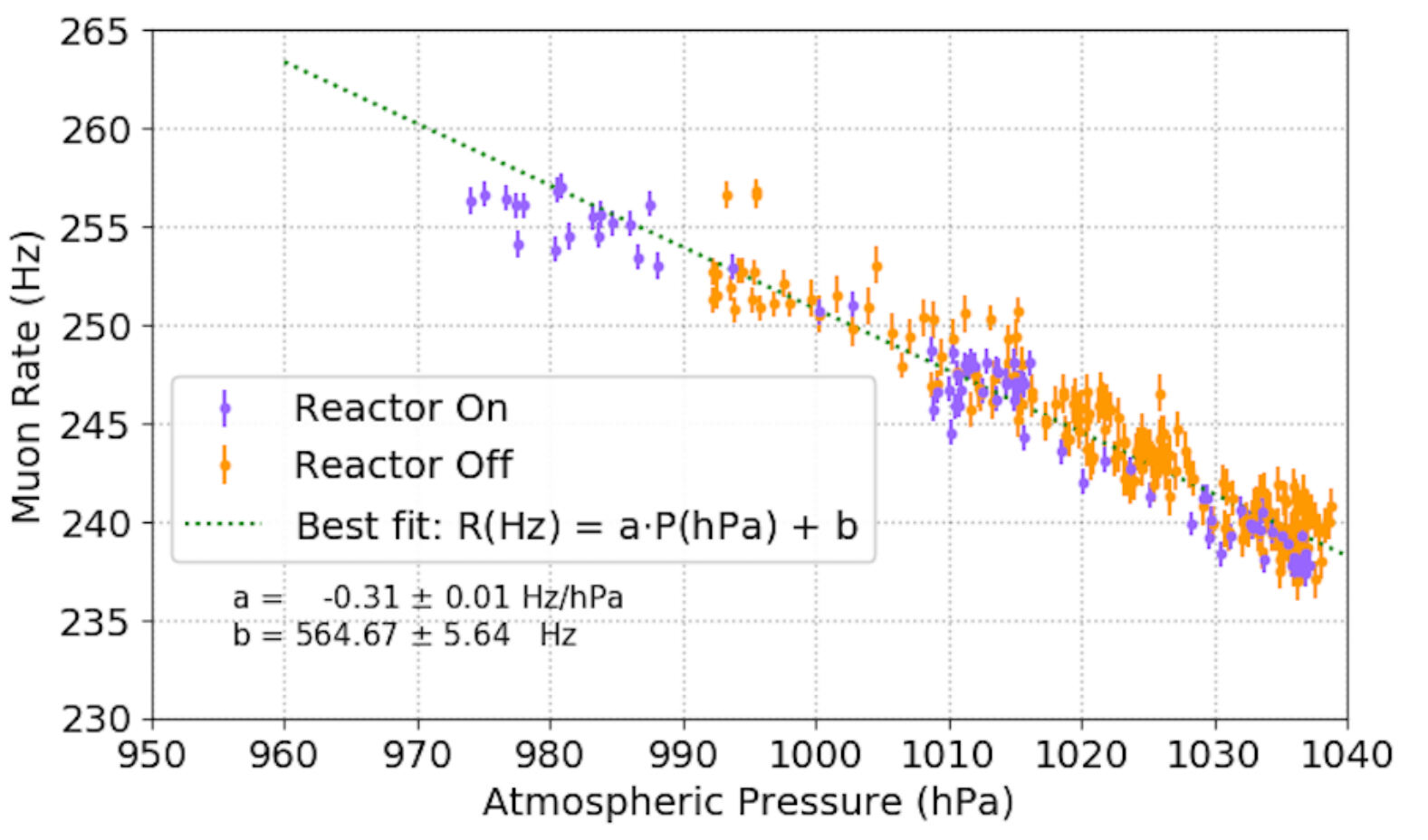}
\includegraphics[width=.315\textwidth]{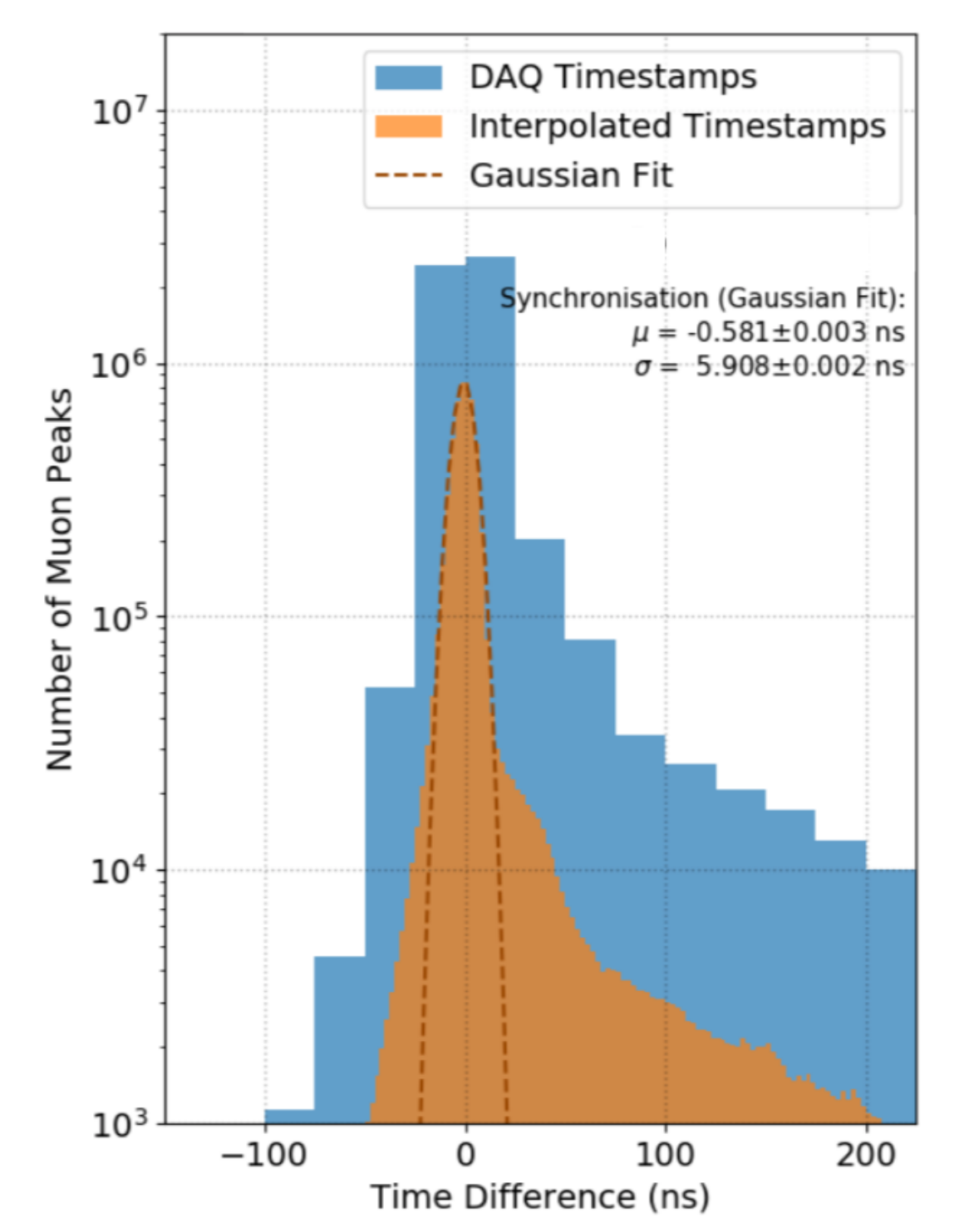}
\vspace*{-0.3cm}
\caption{{\small (Top) Evolution of the reconstructed muon track rate (blue), the atmospheric pressure (red), 
and the outside temperature (brown). (Bottom left) Linear correlation between the reconstructed muon track 
rate and the atmospheric pressure. (Bottom right) Distribution of the time difference between all energy 
deposits originating from the same muon for a large sample of reconstructed muon tracks, in units of the DAQ timestamps in blue, and after interpolation in orange.}}
\label{fig:muons_rate}
\end{figure}

The reconstructed muon track rate, which is about 250\,Hz, can be used as a standard observable, providing uniformity maps of the detector response and an effective tool to control the stability over time. As expected, we observe a linear relationship between the muon rate and the atmospheric pressure (see Fig.\,\ref{fig:muons_rate}). The muon tracking also allows to verify the time synchronisation of the detector channels. 
The time in which a muon crosses the detector is negligible compared to the DAQ sampling time and 
the deposited energy in each cell has to be detected simultaneously. As shown on Fig.\,\ref{fig:muons_rate}, the detection planes are synchronized within 6ns. The origin of the double structure in the positive tail of the time difference distribution is mainly caused by afterpulsing effects when vertical muons deposit a lot of energy, in a single cube. The first shoulder is due to the decay time of a high amplitude pulse back to the pedestal. During this time period, any spurious signal (noise pixel avalanche or afterpulse, is superimposed on a nonzero baseline and can again be reconstructed as an independent energy deposit.~\\

\begin{figure}[h!]
\centering
\includegraphics[width=.9\textwidth]{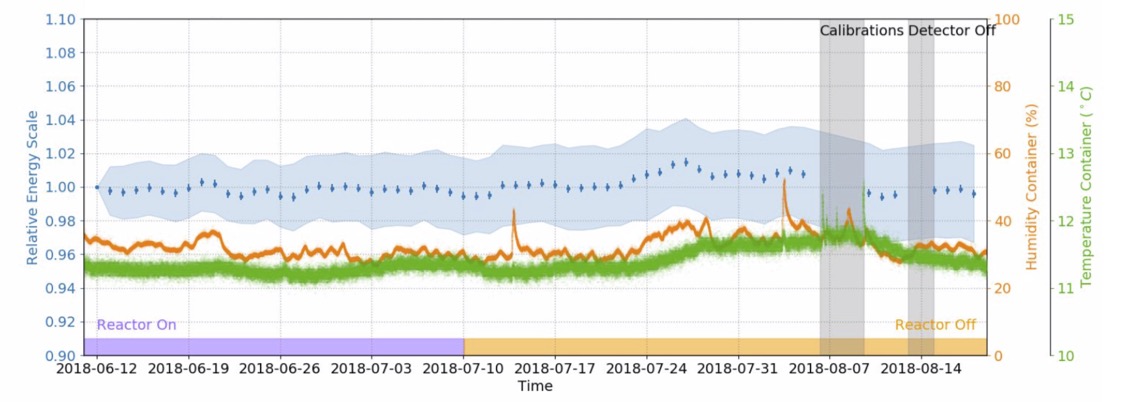}
\vspace*{-0.1cm}
\caption{{\small Evolution of the relative energy scale determined by the muon calibration, corrected for gain and baseline variations, and averaged over all channels. The error bars on the points represent the statistical uncertainty, the blue band represents the total uncertainty (stat. + syst.). The systematic uncertainty is dominated by the uncertainties in the muon path length trough each cell, resulting from the track fit uncertainties. }}
\label{fig:muons_analysis}
\end{figure}

The tracking algorithm also computes the muon path length in each cell by fitting the $dE/dx$ distributions. 
It is then possible to continuously monitor the stability of the detector response during physics mode. As shown 
in Fig.\,\ref{fig:muons_analysis}, the variation of the energy scale is below 2\% over a data taking period of two months. The uncertainties on the energy scale measurements using muon tracks are dominated by the uncertainties on the track fit and the determination of the corresponding path length in each cell. The small drift in energy scale is likely correlated with changes in temperature and humidity of the detector.   

%

\section{Simulation}\label{sec:simulation}

The simulation of the SoLid detector is divided in two parts: one part models the energy 
loss and scattering of particles, including neutrons, in the SoLid detector and the reactor 
hall, while the second stage models the optical system of the detector, including the scintillator 
response, the optical transport, the photon collection by the MPPCs and the electronics response.

 \subsection{\GEANTfour model}\label{sec:geant4_model}

The first part, SoLidSim, is implemented using the \GEANTfour   simulation 
library \cite{Agostinelli:2002hh}. In order to accurately model the scattering of fast neutrons, 
the propagation of cosmic showers through the detector and the creation of spallation 
products in high-Z materials surrounding the detector, a detailed geometry model of 
the detector surroundings is made. This model, as graphically shown in Fig.\,\ref{fig:br2sim}, is based extensively on detailed blueprints of the reactor building and survey measurements 
performed prior to detector installation and includes as main features the majority of the 
concrete and steel structures of the BR2 containment building, including the cylindrical 
containment building inner and outer walls and dome cap, the concrete floors of level 3, 
where the detector is located, level 2 below the detector and levels 4 to 7 situated above 
the detector. Specific features such as staircases, elevator shafts, crane passways, 
and access holes are included as well. Special care is taken to model in detail the reactor 
fuel tank, the water pool and its concrete walls with beam ports including concrete and steel 
plugs, the 20\,cm thick lead shielding wall in between the SoLid detector and the radial 
beam port facing the reactor core. The inclusion of these structures can be switched off in the tracking 
of particles through the detector to save time and computing power for simulations of IBD events 
or background processes occurring inside the detector. The geometry of the detector includes 
besides the sensitive volume of the detector, all HDPE neutron reflectors, all metal structures 
surrounding the sensitive volume, including the electronics housing, the CROSS system, all 
mounting rails, the container insulation and steel walls, the passive water and PE shielding 
surrounding the detector and its support scaffolding.

 \subsection{Readout simulation}\label{sec:RO_sim}

After modelling the energy deposits or the creation of secondary particles in the detector 
and its surroundings with SoLidSim, the energies deposited in the sensitive volume of the 
detector are translated in detected pixel avalanches in the MPPCs connected to fibres 
surrounding the energy deposit. This is modelled in a standalone readout software library, 
ROSim, which is specific to SoLid, and supported by dedicated lab bench measurements of various parts of the optical system, as described in~\cite{Abreu:2018ajc}. For the optical part, it includes the modelling of the 
scintillation photon production in the PVT and in the ZnS scintillator of the neutron detection screens, 
in particular the non-linearity corrections to the energy response using Birks' law,  the loss of 
scintillation photons due to scattering and absorption in the PVT cubes, the neutron detection screens 
and the wrapping material, the attenuation of the wavelength shifted photons in the optical fibres, 
the reflectivity and absorption losses in the mirrored fibre ends. The Birks constant values for 
PVT and ZnS are based on the literature values for these materials and are respectively 
0.15\,mm/MeV and 0.001\,mm/MeV. The reflectivity of the mirrors on the fibre ends is taken 
to be 80\%. The attenuation 
lengths of each fibre have been accurately measured during in-situ calibration 
campaigns with CROSS and vary between 90\,cm and 115\,cm. The current simulation model 
includes the in-situ measured attenuation length of each fibre. The final number of photons 
is fluctuated according to Poisson statistics and distributed exponentially in time with time 
constants corresponding to the PVT and ZnS scintillator decay constants.\\

The number of photons arriving at the MPPCs are tuned, via the relevant parameters in ROSim, to the values measured with 
calibration campaigns using gamma sources. The readout simulation also takes into account the 
measured dark-count rate, which is generated uniformly across the detector at a rate of 
110\,kHz per channel. The probability of cross-talk in a neighbouring pixel, of about 20\%, 
is also considered. After an avalanche is triggered in an MPPC, the pixel is insensitive for 
incoming photons for a short time period. This pixel recovery is modelled by an exponential 
recovery of the pixel bias voltage with a time constant of 24\,ns.\\

\begin{figure}[h!]
	\centering
         \includegraphics[width=.49\textwidth]{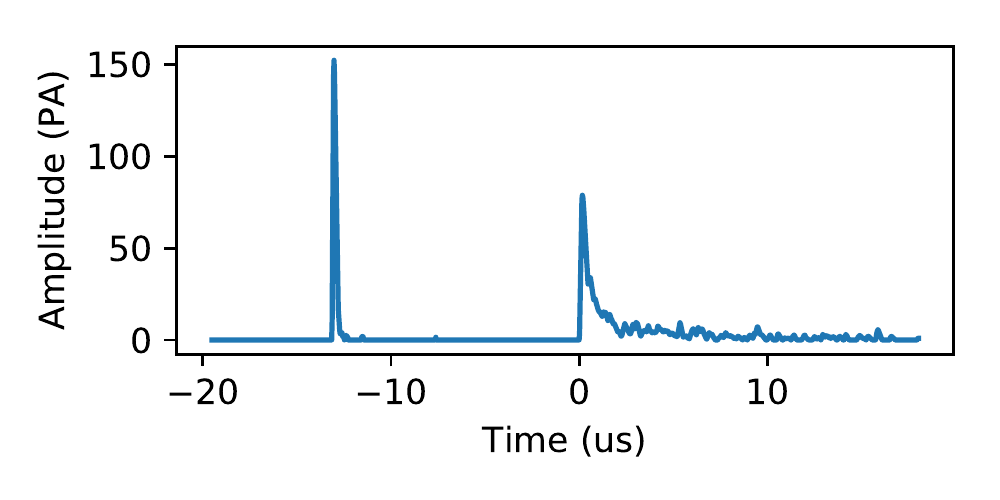}
          \includegraphics[width=.49\textwidth]{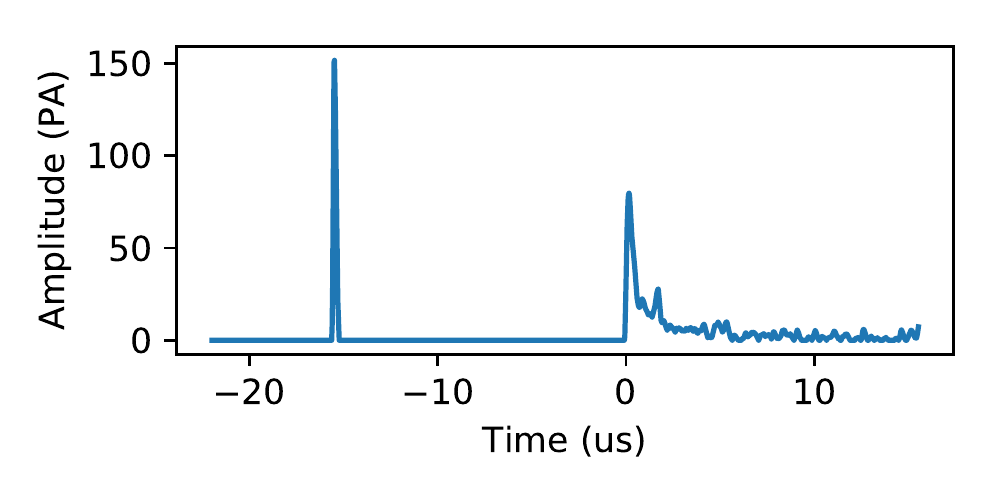}
	\caption{{\small{Examples of waveforms: (Left) IBD candidate reconstructed from data taking during reactor ON period. (Right) IBD event generated with SoLidSim and the readout simulation.}}}
	\label{fig:wf}
\end{figure}

The last stage of the simulation takes into account the shaping and amplification of the MPPC signals, 
and adds a small amount of white noise. This noise is modelled by a Gaussian smearing of the ADC sample amplitudes 
around the nominal baseline. The RMS of the noise corresponds to the values measured in data and 
equals 2 ADC, compared to the amplitude of a single pixel avalanche which is 32 
ADC for an MPPC bias of 1.8\,V above its breakdown voltage. The pulses are finally sampled with 
the same frequency and resolution as the SoLid ADCs and the data is stored in the same format as 
real data for processing by the reconstruction software. An example of an IBD event generated with the 
SoLidSim software and processed by the readout simulation is shown in Fig.\,\ref{fig:wf} and compared 
to an observed IBD candidate. 
Since the IBD trigger operates on themalised neutrons, a proper modeling of the neutron waveforms is essential. We assume that neutrons, once thermalised, are indistinguishable with respect to the process they were produced by. As such, thermalised neutrons from the IBD process, or from a neutron calibration source, should behave identical. Consequently, a trigger bias on a neutron waveform from IBD events should behave the same as one on a neutron waveform from a calibration source. We therefore tune the NS waveforms in the simulation to neutron induced waveforms from data using our neutron calibration sources with a high neutron flux (1000-3800 n/sec), collected by the NS trigger. The true trigger efficiency, as obtained from dedicated neutron calibration campaigns, is applied later in any IBD analysis as an overall scale factor.

%

\section{Data taking and calibration}\label{sec:calibration}

\begin{figure}[h!]
	\centering
	\includegraphics[width=.95\textwidth]{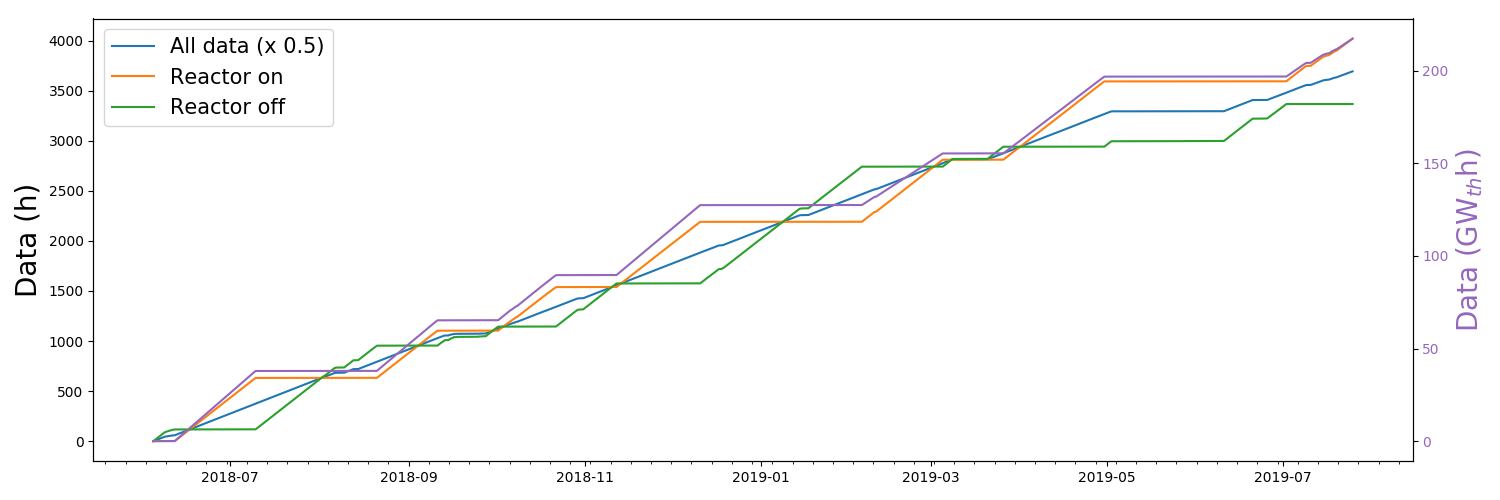}
	\caption{{\small{The integrated data taking time for reactor ON conditions (orange) and reactor OFF background measurements (green) taken over the course of approximately one year of SoLid detector operations. The blue curve shows the nominal physics data taking time, with the exception of source calibrations denoted in Tab.\,\ref{tab:lumi}, and the purple curve shows the integrated BR2 reactor power over time.}}}
	\label{fig:intlumi}
\end{figure}

The SoLid detector was commissioned between February and June in 2018, after which it entered stable physics operations. Since then the experiment has been in continuous operation during all subsequent BR2 reactor cycles and refuelling periods (see section \ref{sec:BR2_description}). We thus collect approximately as much reactor ON data as reactor OFF data. The data taking is guaranteed to last until the end of 2021, but can possibly still be extended if needed,  in order to accumulate enough statistics and achieve a relative measurement almost solely limited by the systematics of the detector.. On a regular basis, in between reactor ON cycles, two to five days are reserved for in-situ detector calibrations using the CROSS system, described in section\,\ref{sec:layout}, with several neutron and gamma sources, as described below. The periods during which the SoLid detector collected physics quality data during reactor ON periods is summarized in Tab.\,\ref{tab:lumi}, and the periods during which calibration data were taken are shown in Tab.\,\ref{tab:cali}. The integrated amount of data taking time under various conditions, together with the integrated BR2 reactor power at which the SoLid detector collected physics data, is shown in Fig.\,\ref{fig:intlumi} over the course of one year of operation.\\

\begin{table}[h!]
	\begin{center}
		\begin{tabular}{c|c|c}
			\textbf{Period} & \textbf{days} & \textbf{Thermal Power (MW$_{th}$)}\\ 
			\hline
			12/06/18 -- 10/07/18 & 28 & 60 \\
			21/08/18 -- 11/09/18 & 21 & 58 \\
			02/10/18 -- 22/10/18 & 20 & 56 \\
			12/11/18 -- 11/12/18 & 29 & 58 \\
			05/02/19 -- 05/03/19 & 28 & 45 \\
			26/03/19 -- 30/04/19 & 35 & 53 \\
			02/07/19 -- 06/08/19 & 35 & 48 \\
			\hline
			Total & 196 & $\langle$ 54 $\rangle$
		\end{tabular}
	\end{center}
			\vspace*{-0.1cm}
	\caption{Data collection periods during BR2 reactor operations and corresponding average thermal power.}
	\label{tab:lumi}
\end{table}

\begin{table}[h!]
	\begin{center}
		\begin{tabular}{c|c|c|c|c}
			\textbf{Year} & \textbf{month} & \textbf{Source} & \textbf{$\langle$ LY $\rangle$ (PA/MeV)} & \textbf{rms LY (PA/MeV)}\\ 
			\hline
			\vspace{0.02cm}
			2018 & August & $^{22}$Na, AmBe & 92.2 & 6.7\\
			2018 & September & $^{22}$Na, $^{207}$Bi, AmBe, $^{252}$Cf & 96.7 & 7.5\\
			2018 & October	& $^{22}$Na, AmBe & 96.2 & 7.4\\
			2018 & December	& $^{22}$Na & 97.0 & 7.4\\
			2019 & January	& $^{22}$Na & 96.0 & 7.5\\
			2019 & May	& $^{22}$Na, AmBe, $^{252}$Cf & 94.5 & 7.3
		\end{tabular}
	\end{center}
	\vspace*{-0.2cm}
	\caption{Calibration periods with neutron and gamma sources during reactor off times, together with the evolution of the mean and rms of the light yields (LY) of all detection cells, as determined by the methods described in section~\ref{sec:E_scale}.}
			\label{tab:cali}
\end{table}
%

 \subsection{Neutron calibration}\label{sec:neutron}

The neutron detection efficiency drives directly the IBD detection efficiency. In SoLid, we aim to perform an absolute flux measurement as well as an oscillation analysis based on the relative flux distortion across the detector. To this end, we aim to determine the neutron detection efficiency per module with an accuracy of around 5\% in absolute and at the \% level in relative. Due to the fact that the coverage of the neutron sources are not uniform within the detector, the calibration is done individually for each of the 12800 detection cells before being averaged.\\

The neutron detection efficiency, hereafter denoted as $\epsilon_n$, can be expressed as the product $\epsilon_n = \epsilon_{capt} \times \epsilon_{reco}$. In the first place, the probability that the neutron gets captured on the neutron detection screens, denoted $\epsilon_{capt}$, depends on the position and initial energy of the neutron, as well as its transport in the experimental set-up. These neutron capture efficiency is determined at the detection cell level for each neutron source and at each source positions, by using dedicated \GEANTfour Monte-Carlo simulations. Secondly, the reconstruction efficiency, denoted, $\epsilon_{reco}$, is the convolution of the probability that the neutron signal trigger the DAQ with the probability that the associated {\em NS} trigger event pass the offline analysis cuts. It does not depend on the neutron origin and is determined by comparing calibration measurements with the dedicated \GEANTfour Monte-Carlo simulations.\\

\begin{figure}[h!]
\centering
\includegraphics[width=.7\textwidth]{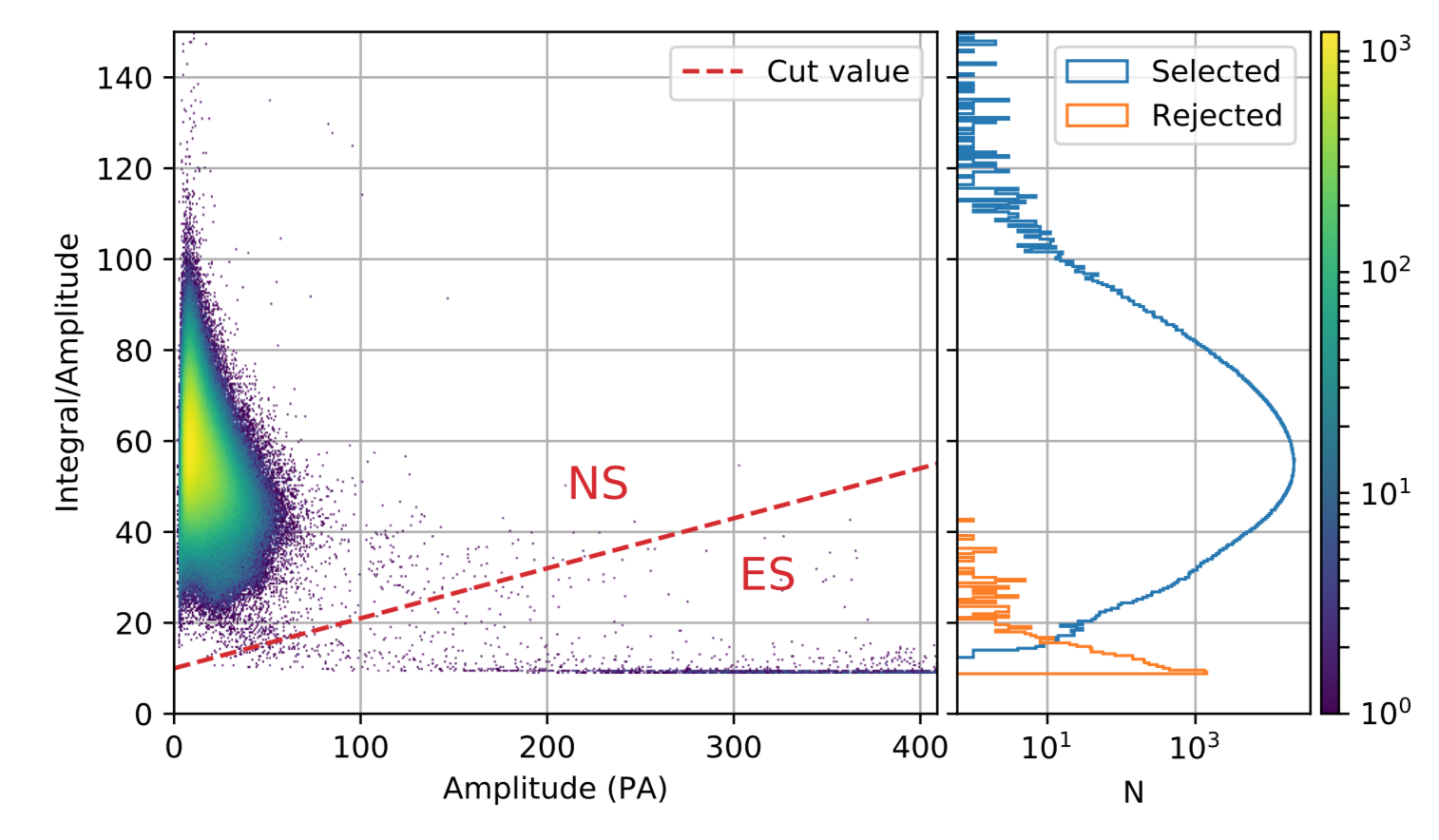}
\caption{{\small{Integral over amplitude ratio versus amplitude, for {\em NS} trigger events, after removal of muon background events. The red dashed line shows the selection requirement used for the particle identification. The right panel presents the projection on the integral over amplitude axis for selected {\em NS} events and rejected  {\em ES} events}}}
\label{fig:NS_calib1}
\end{figure}

The {\em NS} trigger was optimized during the detector commissioning to ensure the largest possible neutron trigger efficiency, while keeping the data rate sustainable (see section\,\ref{sec:daq}). However, this come with a relatively low {\em NS} event purity of about 20\%. The first step consists in removing the muons contribution (see section \ref{seq:SDQM}). In order to reject remaining background, the second requirement is based on an offline pulse shape discrimination using the integral over amplitude ratio. The results are displayed in Fig.\,\ref{fig:NS_calib1}. For the standard data taking in physics mode at BR2, the {\em NS} signals, whose rate does not depend on the reactor operation, can be well separated from the tagged {\em ES} events, with a purity above 99\% after the selection requirements are applied.\\

\begin{table}[h!]
	\begin{center}
		\begin{tabular}{c|c|c|c|c}
			\textbf{Neutron Source} & \textbf{<$E_n$> [MeV]} & \textbf{Initial Activity [n/s]}  & \textbf{Multiplicity} & ~~\textbf{Process}~~\\ 
			\hline
			\vspace{0.02cm}
			AmBe       & 4.16     & 1794 (35) & 1         & {$\alpha + ^9 \rm Be \rightarrow ^{12} \rm C + n $}\\
			$^{252} \rm Cf$ & 2.13     & 3804 (34) & 3.764 (2) & spontaneous fission
		\end{tabular}
		\end{center}
	\vspace*{-0.5cm}
	\caption{Main characteristics of the neutron sources used during calibration runs: mean energy [Mev], initial activity [n/s] and their uncertainties (k=2), neutron multiplicity and physics process.}
\label{tab:neutronsource}
\end{table}

\begin{figure}[h!]
\centering
\includegraphics[width=.7\textwidth]{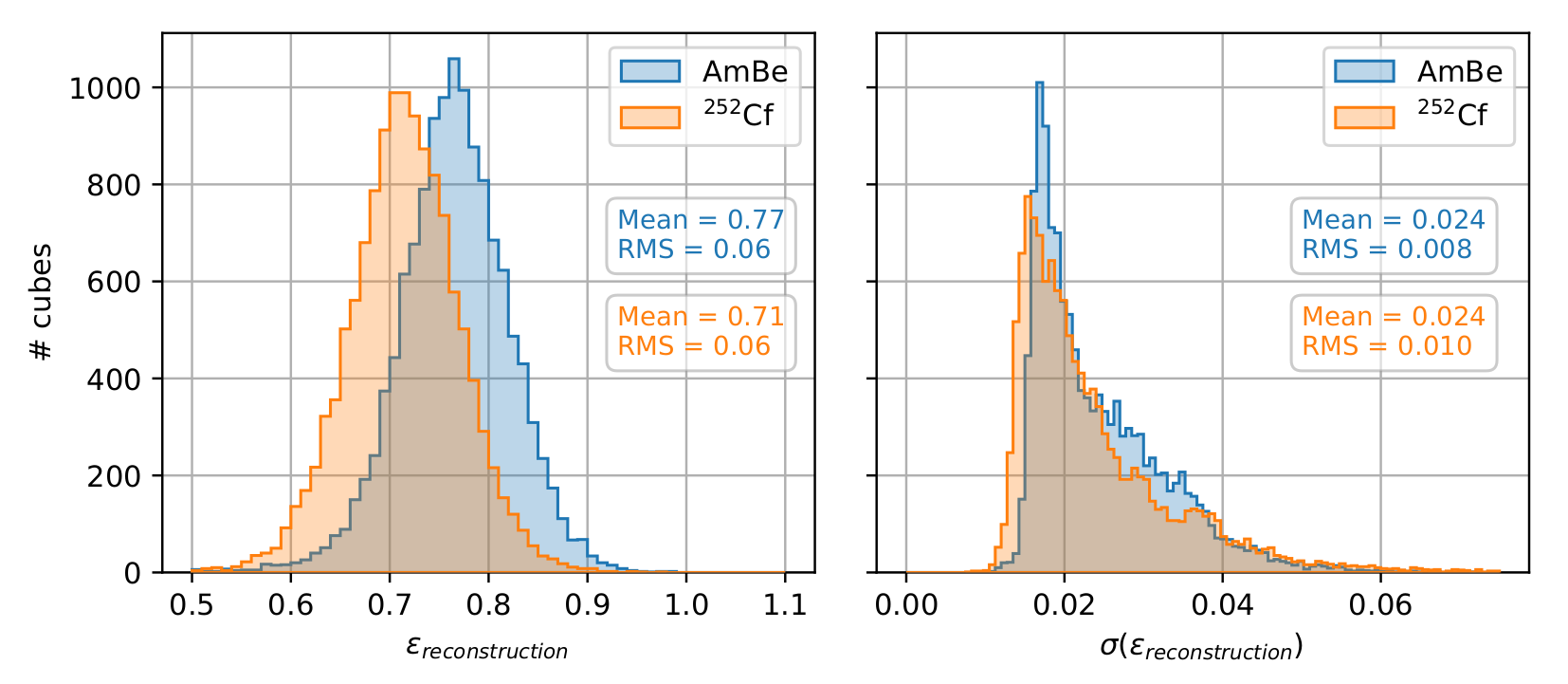}	\vspace*{-0.3cm}
\caption{{\small{ (Left) Neutron reconstruction efficiency for the 12800 cells obtained with the AmBe (blue) and $^{252}$Cf source (orange). (Right) The total uncertainty on the neutron reconstruction efficiency for all detection cells obtained with the two sources (stat. + syst.). The right hand plot shows that the uncertainty is not the same for each cell, which is a consequence of statistics and accessibility during calibration with neutron sources.}}}
\label{fig:NS_calib2}
\end{figure}

During calibration runs, two neutron sources are used, AmBe and $^{252}$Cf, for which activities have been calibrated at the 2\% precision level at the National Physical Laboratory (UK).
The use of two neutron sources, having different characteristics in terms of multiplicity and average neutron energy (see Tab.\,\ref{tab:neutronsource}), allows to estimate the systematics uncertainties related to the Monte-Carlo neutron transport as well as the {\em NS} reconstruction analysis \cite{ref:Pestel_phd}. The two neutron calibration sources are positioned according to the 54 points predefined by CROSS (see section \ref{sec:cross}), with an exposure time of 50 minutes per point for the AmBe source, respectively 60 minutes for the $^{252}$Cf source. Thus, we obtain a cumulative statistic greater than $10^8$  reconstructed {\em NS} events with the AmBe source, respectively $1.7~10^8$ reconstructed {\em NS} events with the $^{252}$Cf source. It corresponds to a statistical uncertainty less than 0.2\% at the level of the plane and of about 5\% in the most unfavorable case at the level of the cell. In turn, the neutron reconstruction efficiency is evaluated cell per cell, with a mean total uncertainty (stat. + syst.) of about~2\% (see~Fig.~\ref{fig:NS_calib2}).

\begin{figure}[h!]
\centering
\includegraphics[width=.8\textwidth]{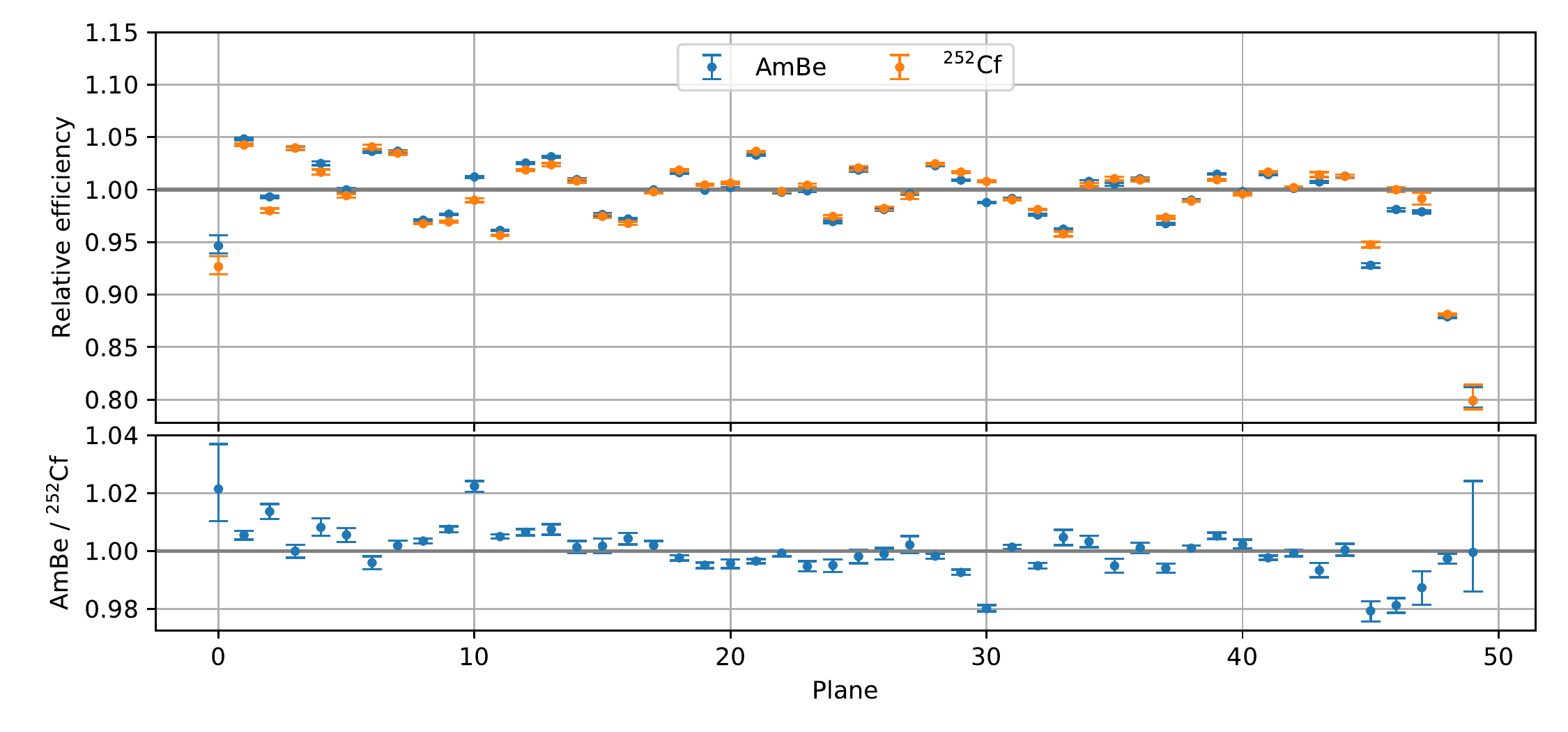}	\vspace*{-0.3cm}
\caption{{\small{ (Top) Relative neutron reconstruction efficiency for the 50 planes obtained with the AmBe (blue) and $^{252}$Cf source (orange). (Bottom) Ratio of the relative neutron reconstruction efficiency obtained with the two different neutrons sources.}}} 
\label{fig:NS_calib3}
\end{figure}

Systematic errors related to the detector dead-time and reconstruction inefficiency are corrected at the plane level. As shown on Fig.\,\ref{fig:NS_calib3}, when cell response are being averaged, the relative neutron reconstruction efficiency per plane is homogeneous at 5\% level across the detector, except for the planes in front and in the back of the detector, due to the higher probability for neutrons to escape. The SoLid detector has an average neutron reconstruction efficiency of 73.9$^{+4.0}_{-3.3}$\%. The absolute systematic uncertainties, which are below 5\% at the module level, are estimated taking into account the difference in efficiency for the two neutron sources and the uncertainty in the activity of those sources. Since the probability of capture of neutrons coming from IBD is of the order of 72\%, we obtain an absolute detection efficiency for IBD neutrons greater than 52\%, with a relative uncertainty between detector modules  below 2\% \cite{ref:Pestel_phd}.
%

 \subsection{Energy scale}\label{sec:E_scale}

To be sensitive to a $\bar \nu_e$ oscillation, the {\em ES} energy reconstruction needs to be 
measured accurately. To that end, the SoLid detector response is calibrated using $\gamma$ sources at various energies (see Tabs.~\ref{tab:gammasource}). The energy scale and its dependence upon the actual deposited energy is currenlty known at the 2\% level.\\

\begin{table}[h!]
	\begin{center}
	\begin{tabular}{c|c|c}
			~~~\textbf{$\gamma$-ray Source} ~~~&~~~~~~\textbf{Energies [MeV]~~(Intensity [\%])}~~~~ &~~~ \textbf{Initial Activity [kBq]}\\ 
			\hline
			\vspace{0.02cm}
			$^{22}$Na       & 0.511 (181)  ; 1.275 (99.9)    &   37      \\
			$^{207}$Bi      & 0.570 (98) ;  1.064 (75) 	; 1.770 (7)   &   37 \\
				AmBe        &   4.43       &  - 
		\end{tabular}
		\end{center}
	\vspace*{-0.5cm}
	\caption{Main characteristics of the $\gamma$-ray sources used during calibration runs: Initial Activity [kBq], $\gamma$-ray energies [MeV] and their respective intensity [\%].}
\label{tab:gammasource}
\end{table}

During standard calibration runs, the energy scale in each detection cell is determined using a 
37 kBq $^{22}$Na gamma source. This source is placed at nine different positions in 
each of the 6 detector gaps, using the CROSS system. To reconstruct the total amount of 
light produced in a given cell, the total number of detected photons originating from the $^{22}$Na decay spectrum per cell must be computed. 
To perform this operation, coincidences are searched for between the two vertical and the two horizontal 
sensors coupled to the four fibres going through each cell. Finally the four amplitudes are summed 
taking into account the gain of each MPPC. Gammas from the $^{22}$Na source (511\,keV and 1270\,keV) interact in the 
PVT mostly through Compton scattering. In addition, given the granularity of the detector planes, 
only a fraction of the total gamma energy is deposited within each PVT cube. Consequently, a broad visible energy spectrum needs to be reconstructed and fitted for one or more Compton edges.\\

\begin{figure}[h!]
\centering
\includegraphics[width=.54\textwidth]{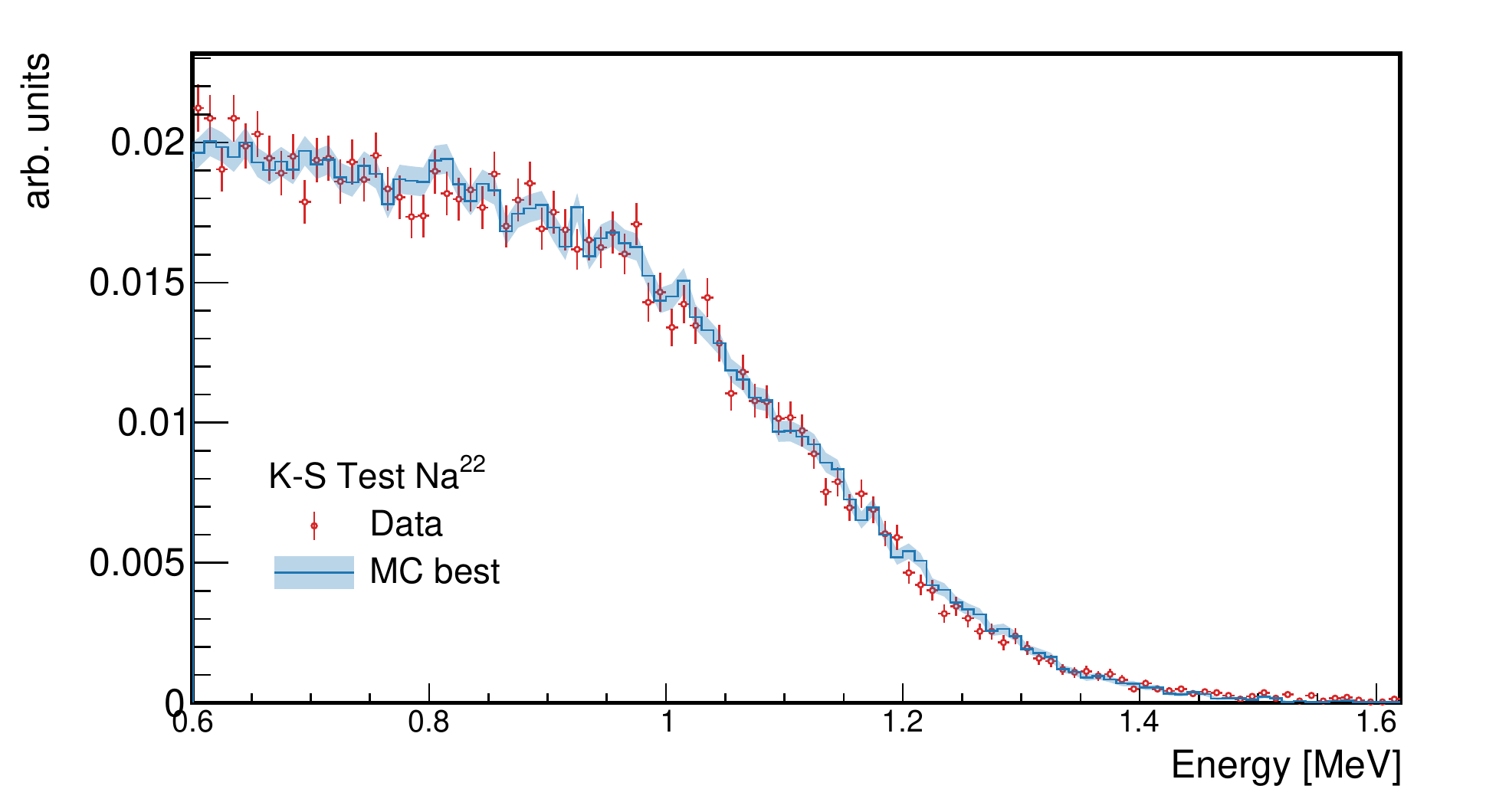}
\includegraphics[width=.45\textwidth]{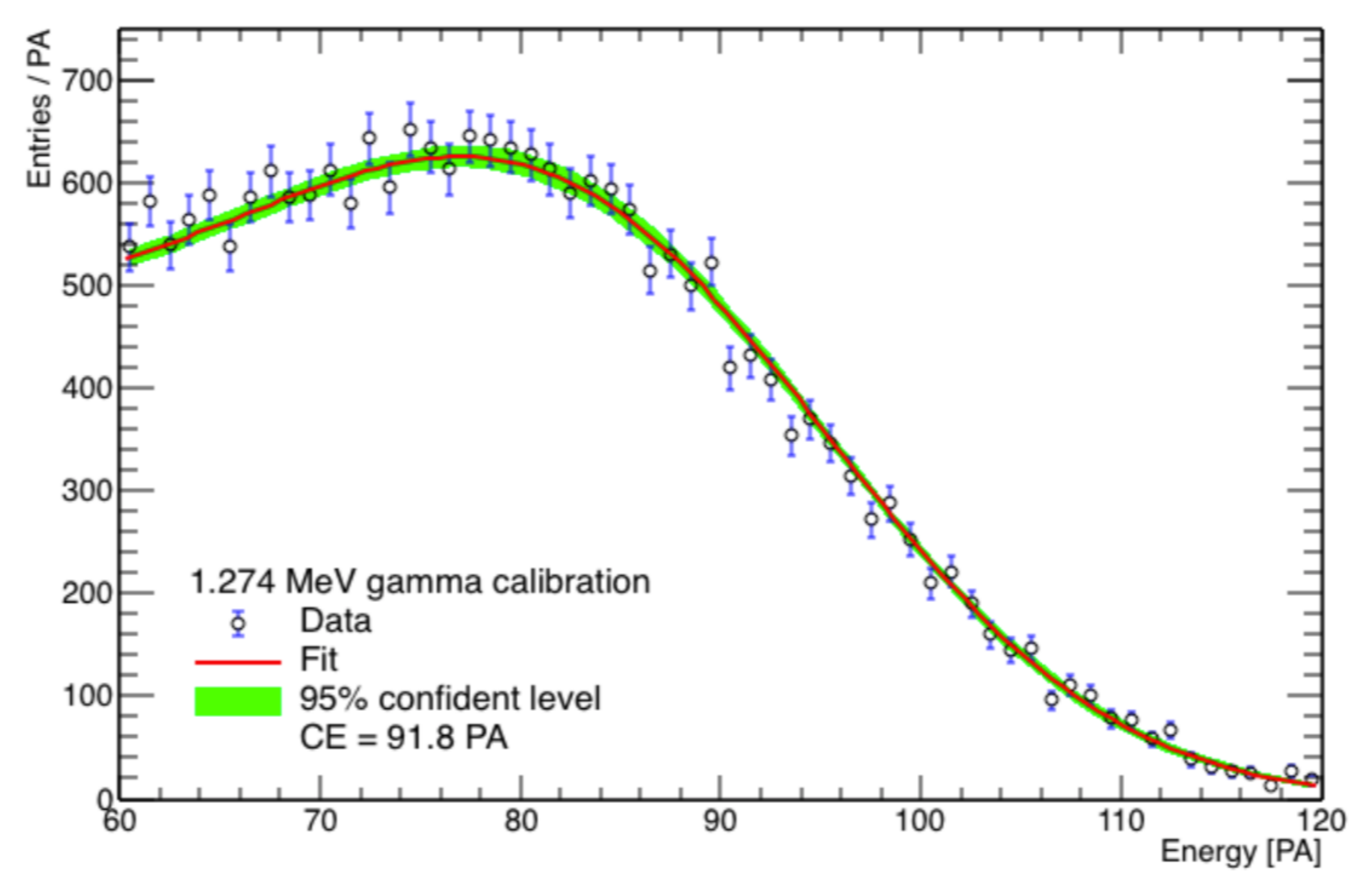}
	\vspace*{-0.3cm}
\caption{{\small{Light yield measurement obtained with $^{22}$Na source by using either a Kolmogorov-Smirnov test (Left) or a fit of the Compton edge profile (Right). The difference in shape 
at low energies is due event selections. The analytical fit only includes events where the cube to be calibrated is the only cube in its plane responsible for triggering the DAQ, therefore prohibiting events in which the 1.2 MeV gamma undergoes a Compton scattering at about 90 degrees. However, at this angle, the deposited energy is about 0.7 MeV. In the Kolmogorov case, this criterion is not imposed.}}}
\label{fig:Lightyield}
\end{figure}

\begin{figure}[h!]
\centering
\includegraphics[width=.96\textwidth]{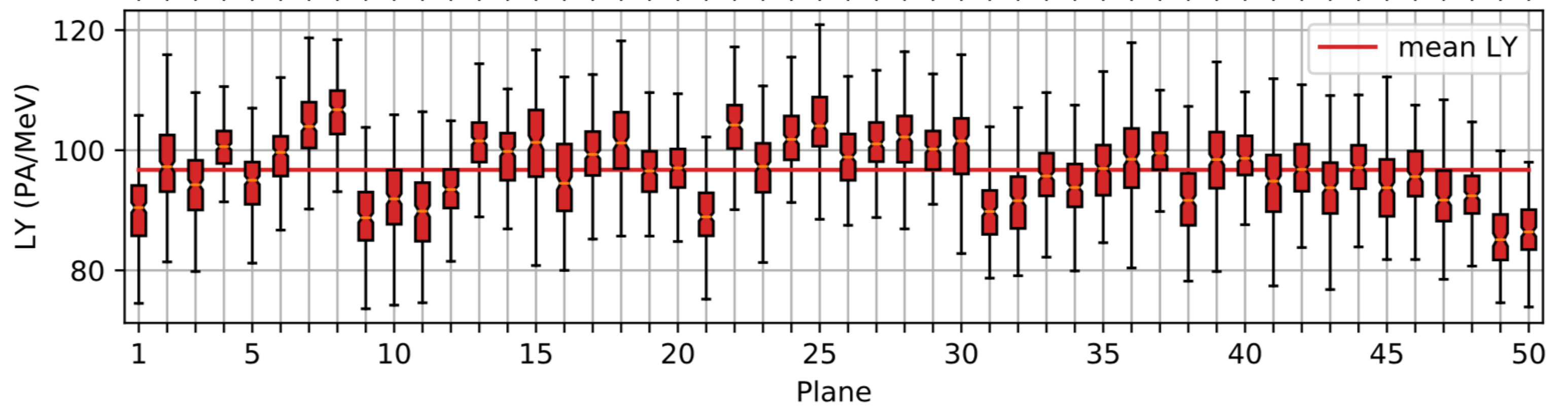}
	\vspace*{-0.3cm}
\caption{{\small{Candle plot for the light yield of the 50 planes obtained with a $^{22}$Na gamma source. The red line represents the mean value over the 50 planes, while the filled boxes represent the cells between 
the first and the third quartiles of each plane (50\% of the data points).}}}
\label{fig:LY_candle}
\end{figure}

\begin{figure}[h!]
	\centering
	\includegraphics[width=.99\textwidth]{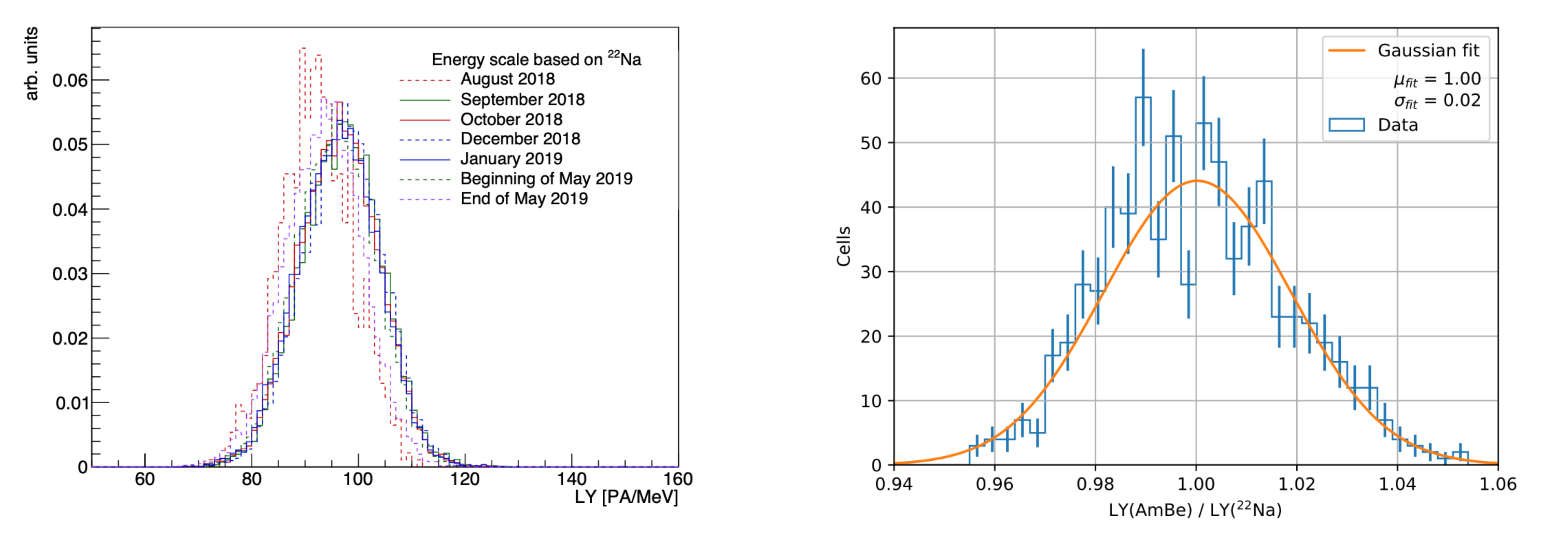}
		\vspace*{-0.3cm}
	\caption{{\small{(Left) Evolution of the energy scale, measured for all cells over time using dedicated in-situ calibration runs with a $^{22}$Na source. (Right) Ratio of the Light Yield (PA/MeV/cell) obtained with $^{22}$Na (1.72\,MeV) and AmBe sources (4.4\,MeV).}}}
	\label{fig:LY_stability}
\end{figure}

\begin{figure}[h!]
	\centering
	\includegraphics[width=.47\textwidth]{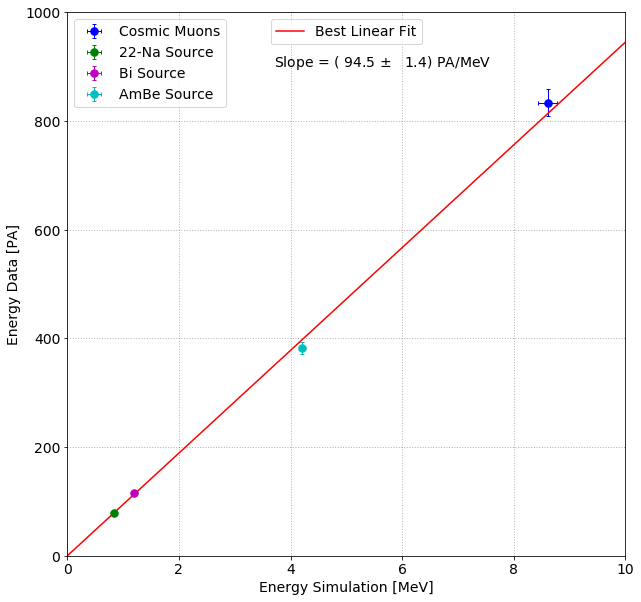}
	\caption{{\small{The reconstructed energy as function of the fitted dE/dx for muons, or the Compton edge position for the gamma calibration sources (abbreviated as Energy Simulation).}}}
	\label{fig:Linearity}
\end{figure}

During the quality assurance process, two methods to tackle the latter issue were developed \cite{Abreu:2018ekw}. The first 
method consists of fitting the Compton edge profile of the spectrum by an analytical function based 
on the Klein-Nishina cross section and the result is compared to the predicted value (see Fig.\,\ref{fig:Lightyield}). 
The second method employs a Kolmogorov-Smirnov test and compares the measured energy spectrum to a 
\GEANTfour simulated sample varying the energy scale and energy resolution (see Fig.\,\ref{fig:Lightyield}). Although the two methods rely on different assumptions and different event selections, the obtained results are consistent within 2\% and meet the required energy scale precision \cite{Abreu:2018ekw}. For the standard data taking in physics mode at BR2, an average of 94 PA/MeV/cell was measured without MPPC cross-talk subtraction, which is estimated to be around 
20\%. The light yield is uniform across the whole detector, as in shown in Fig.\,\ref{fig:LY_candle}, and is stable over time, as can be seen in Fig.\,\ref{fig:LY_stability}. The variation of the mean value of the light yield and the RMS of its distribution are within 2\% over a period of one year. For linearity studies, $^{207}$Bi and AmBe radioactive sources are also used in two detector gaps, in addition to the $^{22}$Na source. The light yield ratio measured with two difference sources is consistent with what is expected for linear behavior as can be seen in Fig.\,\ref{fig:LY_stability}. Figure~\ref{fig:Linearity} shows the reconstructed energy as function of the fitted dE/dx for muons, or the Compton edge position for the gamma calibration sources. It illustrates the linear response of the PVT scintillator over our energy range.\\

%

\section{Conclusion}
The Solid collaboration constructed a 1.6 ton highly segmented neutrino detector based on an affordable dual scintillator technology in the years 2016-2017. 
The use of PVT for calorimetry is cost-effective and provides a linear energy response with an adequate energy resolution of 12\% at 1 MeV, allowing for a very fine spatial segmentation of the fiducial volume.
The 3D segmentation is an intrinsic feature of our detector with the potential to 
reduce intrinsic radioactivity backgrounds, accidentals and multiple recoils induced by fast neutrons. It will eventually also allow us to tag the annihilation gammas from positron interactions, which is a distinct feature of IBD events, provided that the energy reconstruction thresholds can address the low energy deposits per cell of 511~keV gammas.
The detector is capable of operating at very close proximity to a compact research reactor with practically no overburden. Its design is simple and very modular and some of its parameters were improved after a measurement campaign with a single module prototype in 2015. As of the spring of 2018 the full size SoLid detector is in continuous operation at the BR2 research reactor of the SCK\raisebox{-0.8ex}{\scalebox{2.8}{$\cdot$}}CEN in Belgium. The BR2 reactor is operated with highly enriched~$^{235}$U fuel arranged in a very compact geometry, which reduces the uncertainties in the calculation of the incoming electron antineutrino flux and its energy spectrum. The detector has proven to run very stably over long periods of time and can be routinely calibrated with dedicated gamma and neutron sources with an in-situ system. The statistical energy resolution, the energy scale precision and the level of inter-channel response calibration all adhere to or surpass the initial SoLid design specifications. A detailed geometry description and detector response simulation have been developed, allowing for a future validation and understanding of the physical and instrumental backgrounds and an optimisation of the neutrino detection and oscillation measurements.

\section{Acknowledgements}

This work was supported by the following funding agencies: Agence Nationale de la Recherche grant ANR-$16\mathrm{CE}31001803$, Institut Carnot Mines, CNRS/IN2P3 and Region Pays de Loire, France; FWO-Vlaanderen and the Vlaamse Herculesstichting, Belgium; The U.K. groups acknowledge the support of the Science \& Technology Facilities Council (STFC), United Kingdom; We are grateful for the early support given by the sub-department of Particle Physics at Oxford and High Energy Physics at Imperial College London. We thank also our colleagues, the administrative and technical staffs of the SCK\raisebox{-0.8ex}{\scalebox{2.8}{$\cdot$}}CEN for their invaluable support for this project. Individuals have received support from the FWO-Vlaanderen and the Belgian Federal Science Policy Office (BelSpo) under the IUAP network programme; The STFC Rutherford Fellowship program and the European Research Council under the European Union's Horizon 2020 Programme (H2020-CoG)/ERC Grant Agreement \mbox{n. 682474}; Merton College Oxford.

\bibliographystyle{JHEP}
\bibliography{SoLid_detector}

\end{document}